\documentclass{MICE}
\usepackage{lineno}
\usepackage{bibentry}
\nobibliography*
\usepackage{times}
\usepackage{bm}         
\usepackage{amsmath,amssymb,bm,slashed} 
\usepackage{amssymb}    
\usepackage{graphicx}   
\usepackage{verbatim}   
\usepackage{color}      
\usepackage{subfigure}  
\usepackage{hyperref}   
\usepackage{enumerate}
\usepackage{lineno}
\usepackage{textcomp}

\usepackage{fmtcount}   

\usepackage[square,comma,numbers,sort&compress]{natbib}

\def\parenbar{\mathpalette\p@renb@r}
\def\p@renb@r#1#2{\vbox{%
  \ifx#1\scriptscriptstyle \dimen@.7em\dimen@ii.2em\else
  \ifx#1\scriptstyle \dimen@.8em\dimen@ii.25em\else
  \dimen@1em\dimen@ii.4em\fi\fi \offinterlineskip
  \ialign{\hfill##\hfill\cr
    \vbox{\hrule width\dimen@ii}\cr
    \noalign{\vskip-.3ex}%
    \hbox to\dimen@{$\mathchar300\hfil\mathchar301$}\cr
    \noalign{\vskip-.3ex}%
    $#1#2$\cr}}}

\usepackage{amsmath,bm}
\usepackage{tikz}
\usepackage{xcolor}
\usepackage{multirow}
\usepackage{hyperref}

 \numberwithin{equation}{subsection}
 \numberwithin{table}{section}
 \DeclareMathOperator\erf{erf}

\begin{document}
\thispagestyle{empty}

\begin{tabular}{p{0.175\textwidth} p{0.5\textwidth} p{0.225\textwidth}}
  \hspace{-0.8cm}\leftline{}                                 &
  \centering{Muon Ionization Cooling Experiment}                  &
  \rightline{\today} 
\end{tabular}
\vspace{-1.0cm}\\
\rule{\textwidth}{0.43pt}

\renewcommand{\thefootnote}{\ifcase\value{footnote}\or$\dagger$\or*\or
$\ddagger$\or\#\or$\dagger\dagger$\or**\or$\ddagger\ddagger$\or \#\#\or $\infty$\fi}
\begin{center}
  {\bf
    {\LARGE Beam-based detector alignment in the MICE muon beam line} \\
  }
  \vspace{0.2cm}
  Fran\c cois Drielsma\footnote{University of Geneva} \\
  \vspace{-0.0cm}
\end{center}
\renewcommand{\thefootnote}{\arabic{footnote}}
\setcounter{footnote}{0}

\makeatletter

\parindent 10pt
\pagestyle{plain}
\pagenumbering{arabic}                   
\setcounter{page}{1}

\begin{quotation}
\noindent
The Muon Ionization Cooling Experiment (MICE) will perform a detailed study of ionization cooling to evaluate the feasibility of the technique. To carry out this program, MICE requires all of its detectors to reconstruct space points in a globally consistent fashion. The beam-based alignment constants were found to be more accurate than the survey for the scintillating-fibre trackers lodged inside the bores of the superconducting magnets. This alignment algorithm can achieve unbiased measurements of the trackers rotation angles with a resolution of 6\,mrad/$\sqrt{N}$ and of their position with a resolution of 20\,mm/$\sqrt{N}$, with $N$ the number of selected tracks. 
\end{quotation}

\section{Introduction}
\label{sec:introduction}
Intense muon sources are required for a future Neutrino Factory or Muon Collider~\cite{ids_nf, mc}.  At production, muons occupy a large phase-space volume (emittance), which makes them difficult to accelerate and store.  Therefore, the emittance of the muon beams must be reduced, i.e the muons must be ``cooled'', to maximise the muon flux delivered to the accelerator. Conventional cooling techniques applied to muon beams~\cite{beam_cool} would leave too few muons to be accelerated since the muon lifetime is short ($\tau_\mu\sim2.2\,\mu$s). Simulations indicate that the ionization-cooling effect builds quickly enough~\cite{muon_cool_sim} to deliver the flux and emittance required by the Neutrino Factory and the Muon Collider~\cite{nf_iss, mc_design}. The MICE collaboration will study ionization cooling in detail to demonstrate the feasibility of the technique~\cite{mice_proposal}.

\subsection{The Muon Ionization Cooling Experiment}\label{sec:MICE}

\begin{figure}[!htb]
\centering
\includegraphics[width=.875\textwidth]{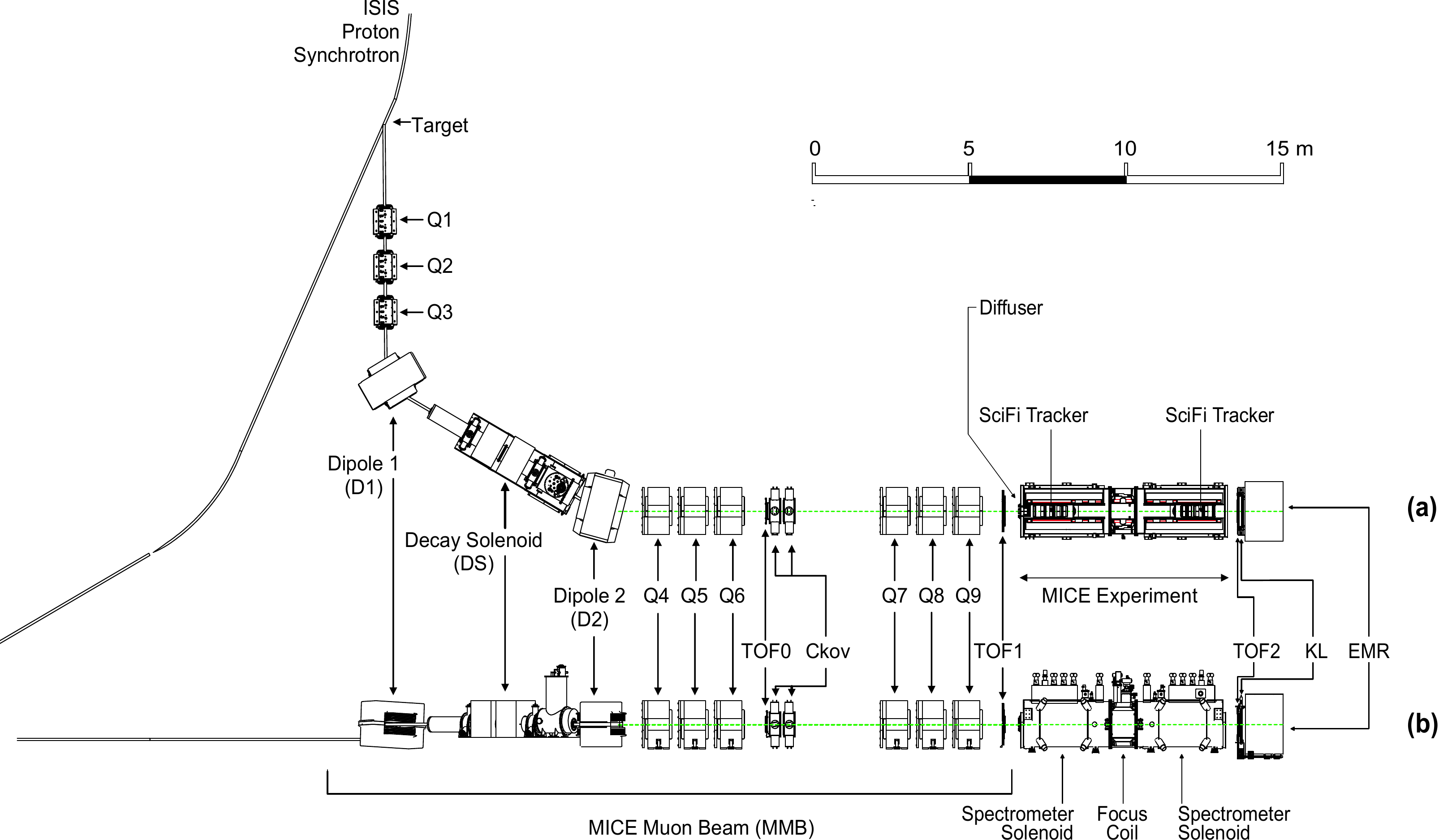}
\caption{(a) Cross-sectional and (b) side views of the MICE Muon Beam Line (MMB) and the MICE experiment. The two scintillating fibre trackers are embedded inside the spectrometer solenoids.}
\label{fig:BeamLine}
\end{figure}

A schematic of the MICE Muon Beam (MMB) and the MICE experiment is shown in figure~\ref{fig:BeamLine} and described in detail in~\cite{mice_bl_inst, mice_demo}. The MMB operates on the ISIS proton synchrotron~\cite{epac04_mice_bl} at the Rutherford Appleton Laboratory. A titanium target~\cite{mice_target} samples the ISIS proton beam, producing pions.  The pions are transported by the upstream quadrupoles, Q1--3, and are momentum-selected at the first dipole, D1. The high field present in the Decay Solenoid (DS) contains the pions to allow them to decay into muons. The second dipole, D2, is used to momentum-select a `muon' beam with high purity or a `pion' beam containing a mix of muons, pions and electrons.  The resultant beam is transported through two quadrupole triplets, Q4--6 and Q7--9, time-of-flight counters~\cite{mice_tofs}, TOF0 and TOF1, and Cherenkov detectors, Ckov~\cite{mice_ckov}, to the cooling cell. The beam emittance is inflated as it passes through the variable-thickness brass and tungsten `diffuser' and is measured in the upstream spectrometer solenoid using a scintillating-fibre tracker. The beam then passes through low-$Z$ absorbers and RF cavities prior to being sampled in the downstream tracker. Upon exiting the cooling cell, the beam is incident upon the final time-of-flight detector, TOF2~\cite{mice_tof2}, a pre-shower detector, KL~\cite{ipac10_mice_pid}, and the Electron-Muon Ranger, EMR~\cite{emr_perf}.

The particle identification system consists of the TOF0, TOF1 and TOF2, Ckov, KL and EMR detectors, which can conceptually be split into two: TOF0--1 and Ckov identifying particle species prior to the cooling cell and TOF2, KL and EMR identifying species after the cooling cell.  In combination with time-of-flight information, the EMR is used to distinguish between muons that have successfully traversed the full cooling cell from those that have decayed en-route.  This aids in the measurement of beam transmission through the cooling cell, as well as reducing the uncertainty on the emittance measurement.

MICE is a single-particle experiment. The emittance reduction is measured by measuring $(x,y,p_x,p_y,p_z)$ of each muon before and after going through the absorber. A sufficient amount of tracks are accumulated during data taking and a sample is assembled during the analysis process to measure the RMS emittance reduction. The single-particle nature of the experiment requires reliable global track matching throughout, i.e. the ability to associate a trace measured in the upstream tracker with one in the downstream tracker but also with the PID detectors. The detectors must reconstruct space points in a globally consistent fashion to guarantee reliable and efficient track matching.

\subsection{Surveys}
\label{sec:surveys}
The baseline for the beam-based alignment is the surveys of the detectors in the hall using laser telemetry. Surveys were performed regularly throughout the MICE Step IV commissioning phase and data taking period. The TOF1 time-of-flight hodoscope was moved periodically to access the upstream end of the superconducting solenoids and resurveyed systematically. The downstream PID detectors module, composed of TOF2, the KL and the EMR, was also repositioned on occasion. The focus coil module was moved in and out of the beam line to change absorbers. Each of these events was followed by a complete resurvey.

\begin{figure}[!htb]
\begin{minipage}[b]{.49\textwidth}
\centering
\includegraphics[width=\textwidth]{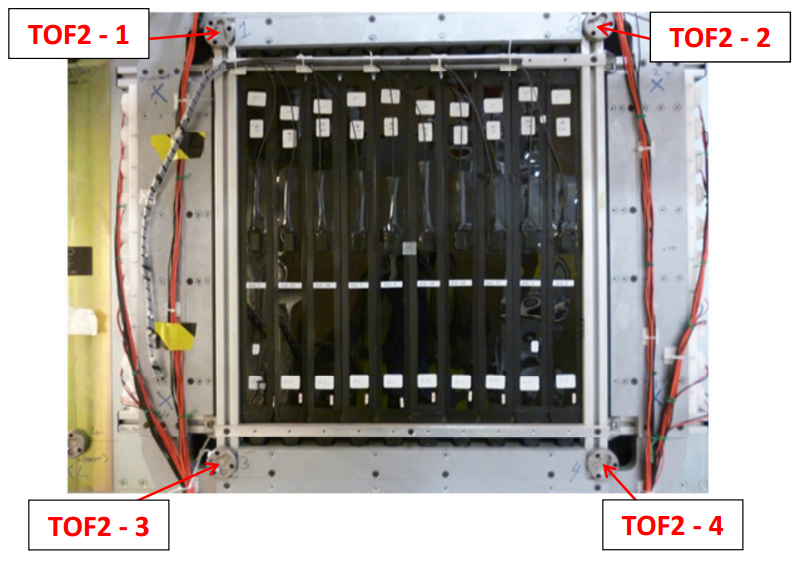}
\caption{Picture of the TOF2 time-of-flight hodoscope and its four survey monuments labelled TOF2.1--2.4.}
\label{fig:tof2_monuments}
\end{minipage}
\hfill
\begin{minipage}[b]{.46\textwidth}
\centering
\includegraphics[width=\textwidth]{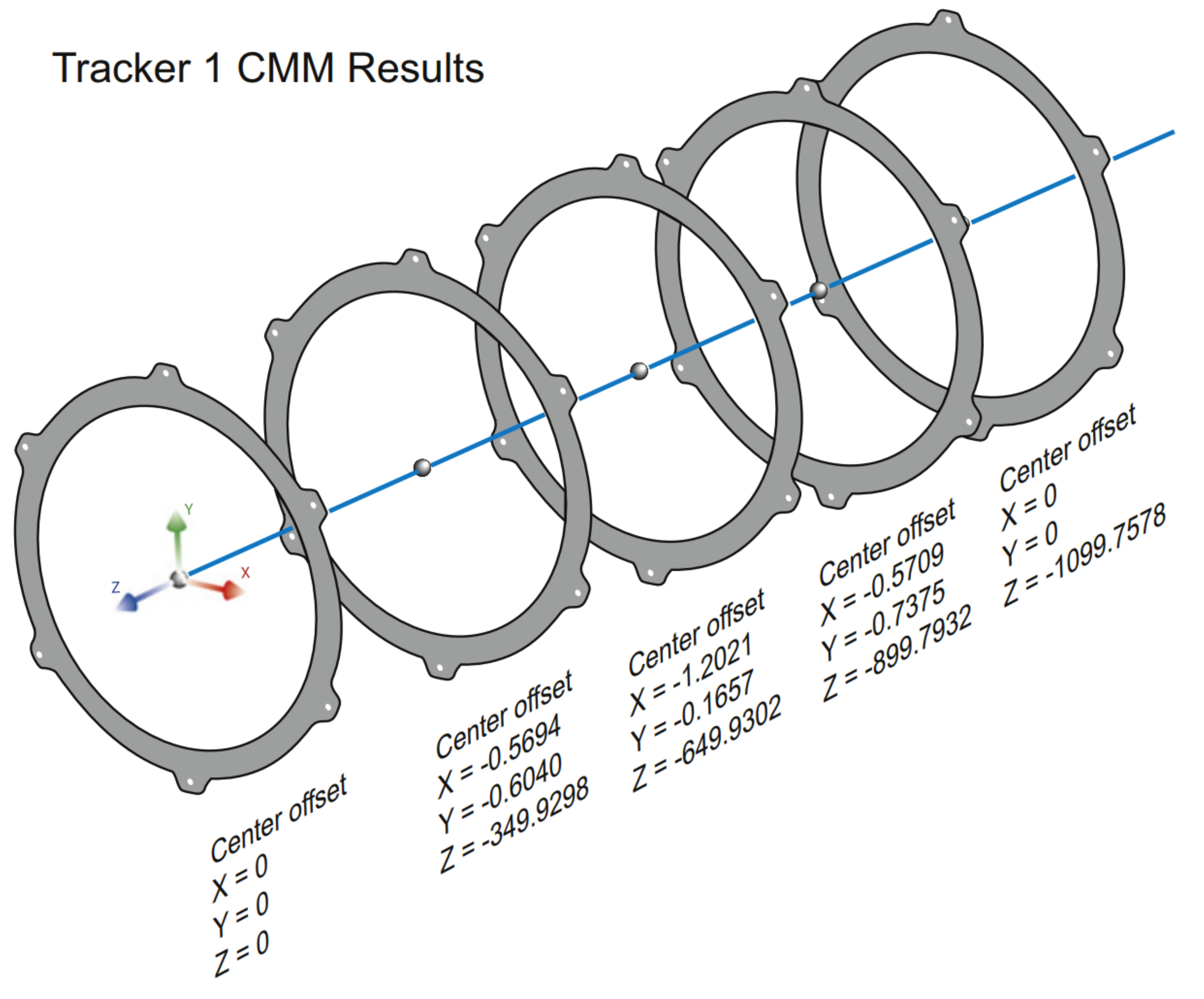}
\caption{Disposition of the downstream tracker stations along with the CMM measurements of their position with respect to the reference axis.}
\label{fig:cmm_measurements}
\end{minipage}
\end{figure}

The PID detectors are each equipped with at least four survey monuments and are surveyed directly~\cite{tof0_survey, tof1_survey, tof2_survey, emrkl_survey}. The two scintillating fibre trackers, nested in the superconducting solenoids, can not be accessed directly. The upstream and downstream flanges of each solenoid are surveyed and the end plate of the trackers are surveyed with respect to the flanges~\cite{ss_survey}. The estimated position of the trackers within the bores are inferred from these measurements. A laser theodolite is used to locate the monuments with respect to the datum point situated under the second dipole magnet, D2. The surveyed positions of the detectors for the 2017/01 ISIS user cycle are summarized in table~\ref{tab:surveys}. Figure~\ref{fig:tof2_monuments} shows a picture of TOF2 and the location of its survey monuments.

\begin{table}[!h]
 \centering
 \begin{tabular}{c|c|c|c|c|c|c}
 & $x_M$\,[mm] & $y_M$\,[mm] & $z_M$\,[mm] & $\alpha_M$\,[mrad] & $\beta_M$\,[mrad] & $\gamma_M$\,[mrad] \\
 \hline
 TOF0 & 1.919 & 1.565 & 5287.247 & 6.030 & 5.252 & 3.718 \\
  TOF1 & -3.738 & -0.913 & 12929.563 & -5.083 & 0.033 & 2.768 \\ 
  TKU & \textbf{0.761} & \textbf{2.732} & 14515.055 & \textbf{2.669} & \textbf{-0.041} & 0.000\\
  TKD & \textbf{-1.027} & \textbf{8.943} & 19398.921 & \textbf{-6.435} & \textbf{1.283} & 0.000 \\
  TOF2 & 13.518 & -10.981 & 21139.375 & 10.735 & 6.699 & -1.406 \\
  KL & 16.616 & -12.855 & 21221.649 & 8.717 & 7.994 & -9.841 \\
  EMR & 35.647 & 7.827 & 21937.889 & -2.527 & 6.857 & -0.118 \\
 \end{tabular}
 \caption{Survey of the detectors in the MICE hall with respect to the datum point at D2 during the 2017/01 ISIS user cycle. The bold figures are the ones sensitive to the beam-based detector alignment.}
 \label{tab:surveys}
\end{table}

Before being placed inside the magnets, each tracker was surveyed independently using a coordinate-measuring machine (CMM). This ensures that the position of the five stations is well known within each tracker with respect to the end plate. Figure~\ref{fig:cmm_measurements} shows the disposition of the stations in the downstream scintillating fibre tracker and their position as measured by the CMM. The reference position is the axis that joins the centre of station 1 to the centre of station 5. The positions of stations 1 to 3 are measured with respect to that axis. The beam can be used to check the tracker station alignment.

Special care is taken during the installation of the trackers within the magnet bores. The installation platform is adjustable to enable the tracker to be aligned with the bore of the solenoid. The tracker sits on four adjustable feet, two at each end. The adjustable feet are used to align the tracker with the magnetic axis of the solenoid. Once this has been done, the location bracket is fitted. The location bracket locks the tracker in its $z$ and azimuthal positions.

\section{Analysis method}
The beam based alignment was initially designed as a consistency check of the detector surveys. The tracking detectors (i.e. the two scintillating fibre trackers and the EMR) are used to extract a straight track fit, propagate the track into other adjacent detectors and compare the predicted position with the measured position. Reliable surveys should yield mean residuals of zero.

It was found that, although the EMR tracks match when propagated to other downstream PID detectors, the tracker tracks do not. A data driven analysis was developed to extract the position and rotations of the two trackers in to the global coordinate system. As the space points reconstructed by PID detectors are globally consistent with their surveyed position, the TOF1 to TOF2 detectors are chosen to construct a reference axis with respect to which to align the trackers.

This section defines the nomenclature used for the alignment constants and describes the measurements required to perform the beam-based alignment. 

\subsection{Module placement}
Each detector is defined as a module in the MICE geometry. A module describes where the elements that compose the detector are placed with respect to the detector centre, i.e. in local coordinates. The placement of the module in global coordinates is entirely defined by the location of its centre $(x_\mathrm{M},y_\mathrm{M},z_\mathrm{M})$ and a set of Tait-Bryan angles $(\alpha,\beta,\gamma)$. The rotation about $x$, $\alpha$, is called pitch, about $y$, $\beta$, is called yaw and about $z$, $\gamma$, is called roll. For a set of local coordinates $(x,y,z)$, the global coordinates are reconstructed as
\begin{equation}
\begin{pmatrix}
\xi \\
\upsilon \\
\zeta  
\end{pmatrix} = \mathcal{R}
\begin{pmatrix}
x \\
y \\
z  
\end{pmatrix} +
\begin{pmatrix}
x_\mathrm{M} \\
y_\mathrm{M} \\
z_\mathrm{M}  
\end{pmatrix}\, ,
\label{eq:transfo}
\end{equation}
with $\mathcal{R}$ the module rotation matrix. In the small angle approximation, which is an excellent working hypothesis considering the surveys, $\sin\theta\sim\theta$ and $\cos\theta\sim1$, which yields a matrix of the form:
\begin{equation}
\mathcal{R} = R_xR_yR_z =
\begin{pmatrix}
1 & 0 & 0 \\
0 & 1 & -\alpha \\
0 & \alpha & 1  
\end{pmatrix}
\begin{pmatrix}
1 & 0 & \beta \\
0 & 1 & 0 \\
-\beta & 0 & 1  
\end{pmatrix}
\begin{pmatrix}
1 & -\gamma & 0 \\
\gamma & 1 & 0 \\
0 & 0 & 1  
\end{pmatrix}
=
\begin{pmatrix}
1 & -\gamma & \beta \\
\gamma+\alpha\beta & 1 & -\alpha \\
-\beta+\alpha\gamma & \alpha-\beta\gamma & 1  
\end{pmatrix}\, .
\end{equation}
This matrix is obtained in the MAUS (MICE Analysis User Software) conventions of performing the rotation about $z$ first, then $y$, then $x$. Getting rid of the second order ($\sim\theta^2$) corrections, the matrix simplifies to
\begin{equation}
\mathcal{R} = 
\begin{pmatrix}
1 & -\gamma & \beta \\
\gamma & 1 & -\alpha \\
-\beta & \alpha & 1  
\end{pmatrix}\,.
\end{equation}
Using this expression of the rotation matrix, equation~\ref{eq:transfo} can be rewritten as
\begin{equation}
\begin{pmatrix}
\xi \\
\upsilon \\
\zeta  
\end{pmatrix} =
\begin{pmatrix}
x-\gamma y+\beta z+x_\mathrm{M} \\
y+\gamma x-\alpha z+y_\mathrm{M} \\
z-\beta x+\alpha y+z_\mathrm{M}
\end{pmatrix}\, .
\label{eq:rot_matrix}
\end{equation}

For each detector, there are six potential unknowns: $(x_\mathrm{M}, y_\mathrm{M}, z_\mathrm{M}, \alpha, \beta, \gamma)$. Some simplifications can be made to lower the amount of unknowns. The $z_M$ coordinate of each detector and the $(x_\mathrm{M}, y_\mathrm{M}, \alpha, \beta, \gamma)$ coordinates of the PID detectors are known to great accuracy from the survey. The scintillating fibre trackers are challenging to survey due to their embedding in the two spectrometer solenoids. The beam-based detector alignment is critical to find the $(x_\mathrm{M}, y_\mathrm{M}, \alpha, \beta, \gamma)$ constants for each tracker.

\subsection{Sought after measurements}
\label{sec:measurements}

The location of the TOFs is known to great accuracy and are used as the reference for the tracker alignment. The line that joins the centre of TOF1 with the centre of TOF2 is chosen to be the reference axis. A deviation from this axis is considered as a misalignment of the trackers.

Multiple scattering in the beam line does not allow to do the alignment on single particle basis but works for a larger sample of particles. The mean residual angles and positions of tracker $t=u,d$ with respect to the TOF12 axis are an essential and powerful tool to infer the correction factors $(x_T, y_T, \alpha_T, \beta_T, \gamma_T)$.

Each TOF provides a single space point in the global coordinate system $(\xi_i, \upsilon_i, \zeta_i)$ with $i$ the ID of the TOF. This position is assumed to be the true position with a large uncertainty due to the limited granularity of the detector ($\sigma_x\sim\sigma_y\sim17$\,mm). The gradients of the track between the two TOFs are reconstructed as:
\begin{equation}
\psi'_{12}=\frac{\psi_2-\psi_1}{\zeta_2-\zeta_1},\,\psi=\xi,\upsilon.
\end{equation}
The extrapolated position of the TOF reference track in the centre of tracker $t=u,d$ is
\begin{equation}
\psi_{12}^t=\psi_1+\frac{\psi_2-\psi_1}{\zeta_2-\zeta_1}(\zeta_T-\zeta_1)=(1-\chi_T)\psi_1+\chi_T\psi_2,\,\psi=\xi,\upsilon,
\end{equation}
with $\chi_\mathrm{T}=(\zeta_T-\zeta_1)/(\zeta_2-\zeta_1)$, the fractional distance from TOF1 to the tracker centre.

Tracker $t=u,d$ samples the particle track in five different stations $(x_t^j, y_t^j, z_t^j)$, with $j=1,\dots,5,$. This allows for the reconstruction of a straight track with gradients $x'_t$ (resp. $y'_t$) in the $xz$ (resp. $yz$) projection and its position at the centre, $(x_t, y_t, 0)$. No assumption is made on the prior position of the tracker and hence the coordinates and gradients are returned in local coordinates, i.e. assuming a tracker perfectly aligned with the beam axis, whose centre lies at $z=0$.

Using the rotation matrix in equation~\ref{eq:rot_matrix}, the position of the track in the global coordinate system at the level of the tracker centre reads
\begin{equation}
\begin{pmatrix}
\xi_t \\
\upsilon_t \\
\zeta_t  
\end{pmatrix} =
\begin{pmatrix}
x_t-\gamma_\mathrm{T} y_t + x_T \\
y_t+\gamma_\mathrm{T} x_t + y_T \\
\beta_T x_t+\alpha_T y_t + z_T
\end{pmatrix}.
\end{equation}

A local increment of $(dx_t,dy_t,dz_t)$ in a tracker translates to global increments of
\begin{equation}
\begin{pmatrix}
d\xi_t \\
d\upsilon_t \\
d\zeta_t 
\end{pmatrix} =
\begin{pmatrix}
dx_t - \gamma_T dy_t +\beta_Tdz_t \\
dy_t + \gamma_T dx_t - \alpha_Tdz_t \\
dz_t - \beta_T dx_t+\alpha_T dy_t
\end{pmatrix}
,
\end{equation}
which in turn correspond to global gradients of
\begin{equation}
  \begin{gathered}
    \xi'_t = \frac{d\xi_t}{d\zeta_t} = \frac{dx_t-\gamma_Tdy_t+\beta_Tdz_t}{dz_t\left(1-\beta_Tx'_t+\alpha_Ty'_t\right)} \simeq x'_t-\gamma_T y'_t +\beta_T, \\
    \upsilon'_t = \frac{d\upsilon_t}{d\zeta_t} = \frac{dy_t+\gamma_Tdx_t-\alpha_Tdz_t}{dz_t\left(1-\beta_Tx'_t+\alpha_Ty'_t\right)} \simeq y'_t+\gamma_T x'_t -\alpha_T.
  \end{gathered}
 \label{eq:global_grad}
\end{equation}

In global coordinates, on average, the track reconstructed between TOF1 and TOF2 should agree with the track reconstructed in either tracker, i.e. the mean residuals should be zero. For example, take the following residual gradient:
\begin{align}
\langle \xi'_t-\xi'_{12}\rangle = 0 & \iff \langle x'_t-\gamma_T y'_t +\beta_T-\xi'_{12}\rangle = 0 \nonumber \\[0.2cm]
& \iff \langle x'_t-\xi'_{12}\rangle = \langle \gamma_T y'_t -\beta_T\rangle \nonumber \\[0.2cm]
& \iff \langle x'_t-\xi'_{12}\rangle = \gamma_T\langle y'_t\rangle - \beta_T.
\label{eq:res_xy}
\end{align}
with $\langle x\rangle$ the average of $x$. Equation~\ref{eq:res_xy} uses the linearity of the mean, i.e. $\langle ax+by\rangle= a\langle x\rangle +b\langle y\rangle$, $a$ and $b$ constants. $\langle x'_t-\xi'_{12}\rangle$ is the mean residual between the local angle measured in the tracker, $x'_t$, and the global angle reconstructed between TOF1 and TOF2, $\xi'_{12}$. The value $\langle y'_t\rangle$ is the mean gradient is the $yz$ projection. Both of the aforementioned means can be measured with a resolution progressing as $1/\sqrt{N}$.

Applying the same reasoning to the other three parameters yields the following system of four equations equations with five unknowns:
\begin{equation}
\left\{
\begin{array}{l}
\langle x'_t-\xi'_{12}\rangle = \gamma_T\langle y'_t\rangle - \beta_T \\[0.2cm]
\langle y'_t-\upsilon'_{12}\rangle = -\gamma_T\langle x'_t\rangle + \alpha_T \\[0.2cm]
\langle x_t-\xi_{12}^t\rangle = \gamma_T\langle y_t\rangle - x_T \\[0.2cm]
\langle y_t-\upsilon_{12}^t\rangle = -\gamma_T\langle x_t\rangle - y_T
\end{array}
\right. .
\label{eq:res}
\end{equation}

The roll dependant terms in the equations have but a small influence on the measurement of the other parameters. The roll of the trackers is constrained -- by the supports they are mounted on -- to values of order 1\,mrad within the magnet bores. The mean gradients and positions after selection are of order 1\,mrad and 10\,mm in the worst settings. The systematic error in angle on $\alpha_T$ and $\beta_T$ from the roll, $\gamma_T$, is $\mathcal{O}(10^{-3})$\,mrad and in position on $x_T$ and $y_T$ is $\mathcal{O}(10^{-2})$\,mm. These values are one order of magnitude smaller than the statistical error.

One way to disentangle the roll, $\gamma_T$, from the yaw, $\beta_T$, in the first equation is to bin out $y'_t$ and measure $\langle x'_t-\xi'_{12}\rangle$ as a function of it. A first order fit of the output distribution yields $\gamma_T$ as its gradient and $-\beta_T$ as its $y$-intercept. The same deconvolution can be performed for the other three equations and provides redundant measurements of the roll. 

An attempt at measuring the roll, $\gamma_T$, using this method was inconclusive. The statistical uncertainty for a single measurement was of order 10\,mrad, one order of magnitude larger than the expected roll itself. The roll was ignored in this analysis due to its negligible influence on global track reconstruction.

\section{Sample selection}
The method described in the previous section assumes that the mean residuals can be measured with great accuracy and, more importantly, are unbiased. A bias in one of the residual distributions inevitably introduces a bias in the measurement of the alignment parameters. The method also assumes isotropic scattering and inefficiencies in the detectors involved.

\subsection{Sampling bias}
\label{sec:sampling}
The main source of bias is the Multiple Coulomb Scattering (MCS) in the material between TOF1 and TOF2. The particles used in the alignment go through 15 planes of scintillating fibre in each tracker, the aluminium windows of the spectrometer solenoids and focus coil absorber module and the helium contained in the bore of the magnets. In the absence of MCS, particles go down a straight path between the two detectors and are sampled in the two trackers in a fully nonstochastic fashion.

MCS introduces straggling and an uncertainty on the exact path taken by a particle between the two TOFs. Assume a perfectly Gaussian sample of particles first measured at a given $z_0$. In one dimension, the true position is distributed as $x_0\sim \mathcal{N}(\overline{x}, \sigma_x^2)$ and the true gradient is distributed as $x'_0\sim\mathcal{N}(\overline{x'}, \sigma_{x'}^2)$. With material at $z_0$, the true distributions at a more downstream position $z_1$ are 
\begin{equation}
  \begin{array}{l}
    x_1\sim\mathcal{N}\left(\overline{x}+\overline{x'}z_{01},\, \sigma_x^2+\sigma_{x'}^2z_{01}^2+\theta_{0,0}^2z_{01}^2\right), \\[0.2cm]
    x'_1\sim\mathcal{N}\left(\overline{x'},\,\sigma_{x'}^2+\theta_{0,0}^2\right),
  \end{array}
    \label{eq:x0}
\end{equation}
with $z_{01}=z_1-z_0$. An expression for the RMS scattering angle at $z_0$, $\theta_{0,0}$, can be found in \cite{pdg}:
\begin{equation}
\theta_{0,0}=\frac{13.6\,\text{MeV}}{\beta pc}\sqrt{dz_0/X_{0,0}}\left[1+0.038\ln(dz_0/X_{0,0})\right],
\label{eq:pdg_scat}
\end{equation}
with $\beta=v/c$ and $p$ respectively the particle velocity and momentum and $dz_0/X_{0,0}$ the thickness of the material encountered at $z_0$ in radiation lengths, $X_{0,0}$.

For each position $z_i$ where the sample encounters material, we account for an RMS scattering angle $\theta_{0,i}$. This yields, at the final measurement point $z_n$,
\begin{equation}
  \begin{array}{l}
    x_n\sim\mathcal{N}\left(\overline{x}+\overline{x'}z_{0n},\, \sigma_x^2+\sigma_{x'}^2z_{0n}^2+\sum_{i=0}^{n-1}\theta_{0,i}^2(z_{i+1}-z_i)^2\right), \\[0.2cm]
    x'_n\sim\mathcal{N}\left(\overline{x'},\,\sigma_{x'}^2+\sum_{i=0}^{n-1}\theta_{0,i}^2\right),
  \end{array}
  \label{eq:xn}
\end{equation}
with $z_{0n}=z_n-z_0$. A non-zero mean gradient, $\overline{x'}$, moves the position distribution by $\overline{x'}z_{0n}$, and a spread in gradient translates in a growing spread in position. MCS smears both distributions but preserves the mean. The RMS effective scattering angle, $\theta_0$, and effective distance, $\Delta z$, are defined as
\begin{equation}
\begin{array}{l}
\theta_0 = \sqrt{\sum_{i=0}^{n-1}\theta_{0,i}^2}, \\
\Delta z = \sqrt{\frac{\sum_{i=0}^{n-1}\theta_{0,i}^2(z_{i+1}-z_i)^2}{\sum_{i=0}^{n-1}\theta_{0,i}^2}}.
\end{array}
\label{eq:effective}
\end{equation}

The residuals between the measured and extrapolated distributions are distributed as
\begin{equation}
 \begin{array}{l}
	  x_n-x_0^n=x_n-(x_0+x'_0z_{0n}))\sim\mathcal{N}\left(0, \theta_{0}^2\Delta z^2\right), \\[0.2cm]
	  x'_n-x'_0\sim\mathcal{N}(0,\theta_0^2),
 \end{array}
 \label{eq:residual_dist}
\end{equation}
with $x_0^n$ the track at $z_0$ propagated to $z_n$. The only source of uncertainty on these residuals is the MCS.

Assume that both distributions are measured at $z_0$ and $z_n$. $x_0$ is measured correctly but $x_n$ has a non zero misalignment $\delta x_n$. The mean residual between the two measurements reads
\begin{equation}
    \langle x_n-\delta x_n-x_{n0}\rangle = \langle x_n-(x_0+x'_0(z_n-z_0))\rangle - \delta x_n= - \delta x_n. \\
    \label{eq:true_mean}
\end{equation}

True distributions are inherently unbiased in the measurement of $\delta x_n$. The problem comes from the sampling of the distribution. A tracker station fiducial surface is circular and about 30\,cm in diameter. A typical sample in the MICE beam has a spread of $\sim80\,$mm in the upstream tracker and $\sim200\,$mm in the downstream tracker. The sample is not contained within the trackers and hence the sample mean is not the true mean.

\begin{figure}[!htb]
 \centering
 \includegraphics[width=.9\textwidth]{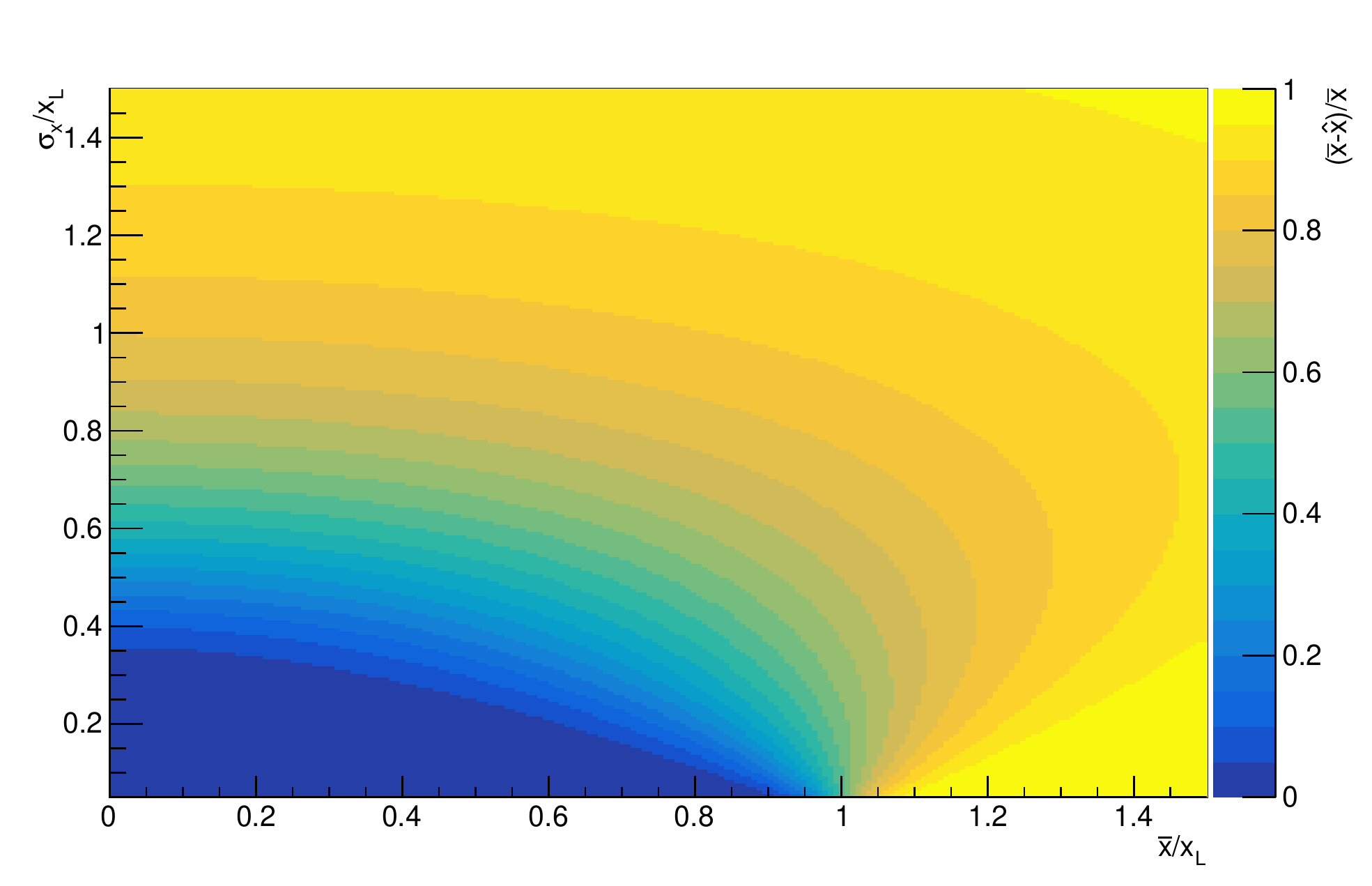}
 \caption{Fractional discrepancy between the sample mean, $\hat{x}$, and the true mean, $\overline{x}$, as a function of the mean and width of the true distribution in units of the range half-width, $x_L$.}
 \label{fig:sample_mean}
\end{figure}

With limits of the sampling defined as $[-x_L,x_L]$, the sample mean of a Gaussian distribution centred around $\overline{x}$ and with a spread $\sigma_x$ reads:
\begin{eqnarray}
\hat{x} & = & \int_{-x_L}^{x_L}x\times\mathcal{N}\left(\overline{x},\, \sigma_{x}^2\right)\mathrm{d}x \nonumber \\
& = & \int_{-x_L}^{x_L}\frac{x}{\sqrt{2\pi}\sigma_{x}}\exp\left[\frac{-1}{2}\frac{\left(x-\overline{x}\right)^2}{\sigma_{x}^2}\right]\mathrm{d}x \nonumber \\
& = & \int_{\frac{-x_L-\overline{x}}{\sqrt{2}\sigma_{x}}}^{\frac{x_L-\overline{x}}{\sqrt{2}\sigma_{x}}}\frac{\sqrt{2}\sigma_{x}z+\overline{x}}{\sqrt{\pi}}e^{-z^2}\mathrm{d}z \nonumber \\
& = & \sqrt{\frac{2}{\pi}}\sigma_{x}\int_{\frac{-x_L-\overline{x}}{\sqrt{2}\sigma_{x}}}^{\frac{x_L-\overline{x}}{\sqrt{2}\sigma_{x}}} ze^{-z^2}\mathrm{d}z + \frac{\overline{x}}{2}\left(\erf\left[\frac{x_L-\overline{x}}{\sqrt{2}\sigma_{x}}\right]-\erf\left[\frac{-x_L-\overline{x}}{\sqrt{2}\sigma_{x}}\right]\right) \nonumber \\
& = & \frac{\overline{x}}{2}\left(\erf\left[\frac{x_L-\overline{x}}{\sqrt{2}\sigma_{x}}\right]-\erf\left[\frac{-x_L-\overline{x}}{\sqrt{2}\sigma_{x}}\right]\right) - \sqrt{\frac{2}{\pi}}\sigma_{x}\exp\left[-\frac{x_L^2+\overline{x}^2}{2\sigma_{x}^2}\right]\sinh\left[\frac{x_L\overline{x}}{\sigma_{x}^2}\right].
\label{eq:sample_mean}
\end{eqnarray}
The relative deviation of the sample mean, $\hat{x}$, in equation~\ref{eq:sample_mean} from the real distribution mean, $\overline{x}$, is drawn in figure~\ref{fig:sample_mean} as a function of the relative true mean and the relative true width. The sample mean preserves the true mean as long as the beam is well contained within the detector fiducial.

At different limits, the formula behaves as expected:
\begin{eqnarray}
 \begin{array}{l}
 \lim\limits_{\overline{x}\rightarrow 0}\hat{x}  =  0 = \overline{x}, \\
 \lim\limits_{\frac{x_L}{\sigma_{x}}\rightarrow +\infty}\hat{x} = \overline{x}\erf(+\infty) -\sqrt{\frac{2}{\pi}}\sigma_{x}\lim\limits_{z\rightarrow +\infty}\left[e^{-z^2}\sinh(z)\right] = \overline{x}, \\
 \lim\limits_{\frac{x_L}{\sigma_{x}}\rightarrow 0}\hat{x} =  \frac{\overline{x}}{2}\left(\erf\left[\frac{-\overline{x}}{\sqrt{2}\sigma_{x}}\right]-\erf\left[\frac{-\overline{x}}{\sqrt{2}\sigma_{x}}\right]\right) = 0.
 \end{array}
\end{eqnarray}
For a distribution centred at 0, the sample mean converges to the true mean. A distribution contained within the detector, i.e. for which $\sigma_{x}/x_L\ll 1$, has a sample mean identical to the true mean as well. The sample mean of a broad distribution, i.e. for which $\sigma_{x}/x_L\gg 1$, is asymptotically zero and thus heavily biased.

Given a significantly off-centre mean of the distribution at $z_n$, the distributions in equation~\ref{eq:residual_dist} are asymmetrically sampled. Given a positive mean, $\overline{x_n}$, a particle that scatters towards the positive $x$ is more likely to scatter out of the detector's fiducial volume. The probability density function of the effective scattering angle for a particle that is expected to hit $x_n$ at $z_n$ is
\begin{equation}
f_{x_n}(\theta) = \frac{C_n}{\sqrt{2\pi}\theta_0}\exp\left[-\frac12\frac{\theta^2}{\theta_0^2}\right]\bm{1}_{[-x_L,x_L]}\left(x_n+\Delta z\theta\right),
\label{eq:fxn}
\end{equation}

\begin{itemize}
\item $\Delta z$, the effective distance between the two measuring stations;
\item $\theta \simeq \frac{x_0^n-x_n}{\Delta z}$, the effective scattering angle;
\item $\theta_0$, the RMS effective scattering angle as described in equation~\ref{eq:pdg_scat};
\item $\bm{1}_{[a,b]}(\cdot)$, the indicator function that takes value 1 for all elements of $[a,b]$ and 0 otherwise;
\item $C_n$, the normalization constant so that $\int_{-\infty}^{+\infty} f_{x_n}(\theta)\mathrm{d}\theta = 1$.
\end{itemize}
The indicator function symbolizes that, for a scattering angle above threshold, the particle scatters out of the fiducial and does not contribute to the PDF. The normalisation constant simplifies to
\begin{eqnarray}
C_n & = & 2\left(\erf\left[\frac{x_L-x_n}{\sqrt{2}\theta_0\Delta z}\right]+\erf\left[\frac{x_L+x_n}{\sqrt{2}\theta_0\Delta z}\right]\right)^{-1} \nonumber \\
& \simeq & 2\left(\tanh\left[\frac{x_L-x_n}{\sqrt{2}\theta_0\Delta z}\right]+\tanh\left[\frac{x_L+x_n}{\sqrt{2}\theta_0\Delta z}\right]\right)^{-1} \nonumber \\
& \simeq & 1+\cosh\left[\frac{\sqrt{2}x_n}{\theta_0\Delta z}\right]/\sinh\left[\frac{\sqrt{2}x_L}{\theta_0\Delta z}\right],
\label{eq:cn}
\end{eqnarray}
which converges to unity if $x_L/(\theta_0\Delta_z)\rightarrow+\infty$ or $x_n\rightarrow0$. The last form in equation~\ref{eq:cn} is obtained using the approximation $\erf(z)\simeq\tanh(z)$.

\begin{figure}[!htb]
	\begin{minipage}[b]{.475\textwidth}
		\centering
		\includegraphics[width=\textwidth]{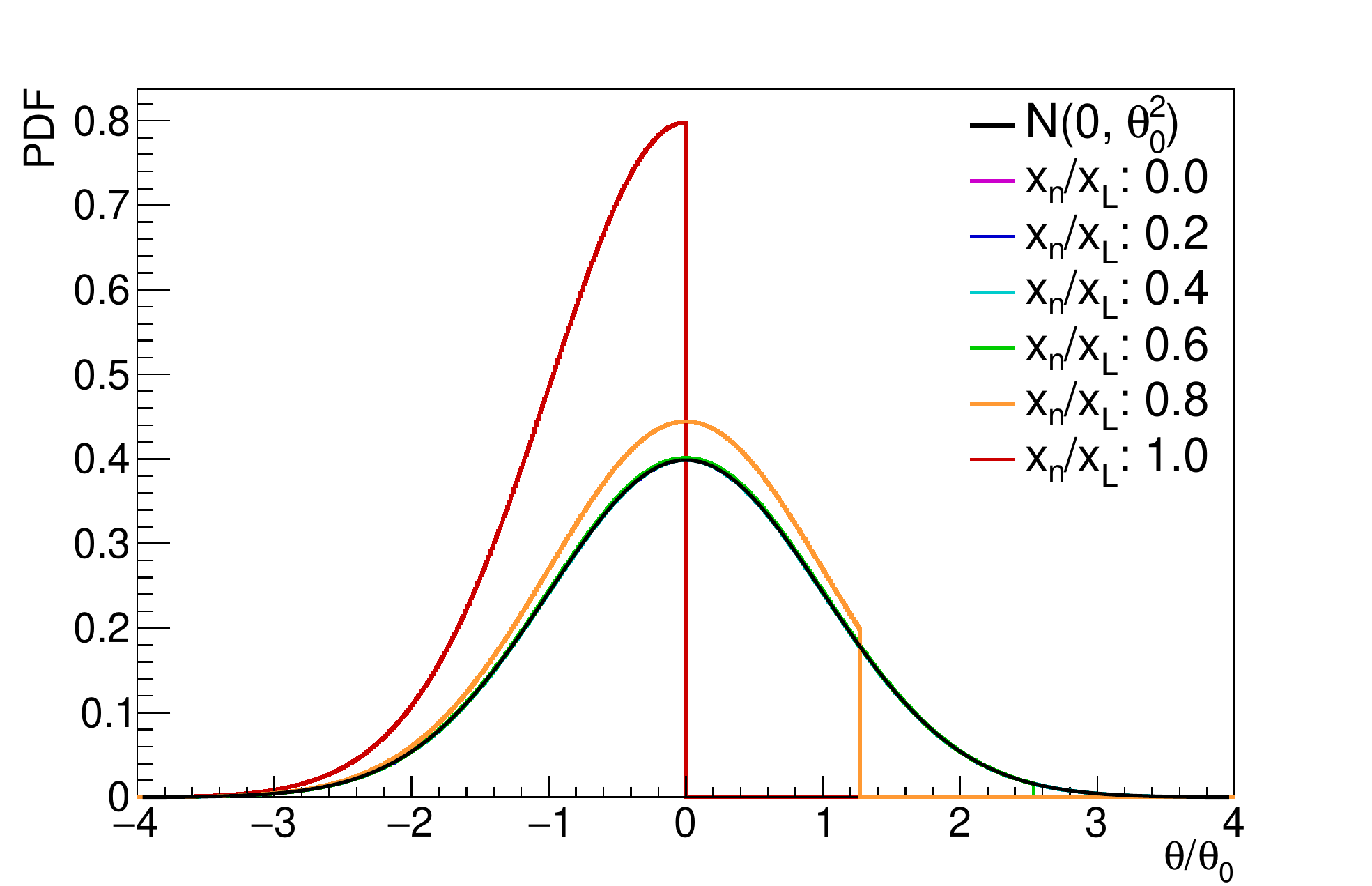}
		\caption{Part of the total scattering angle distribution (black) that does not scrape out for different extrapolated position at the n$^{\mathrm{th}}$ station, $x_n$.}
		\label{fig:angle_xn}
	\end{minipage}
	\hfill
	\begin{minipage}[b]{.475\textwidth}
		\centering
		\includegraphics[width=\textwidth]{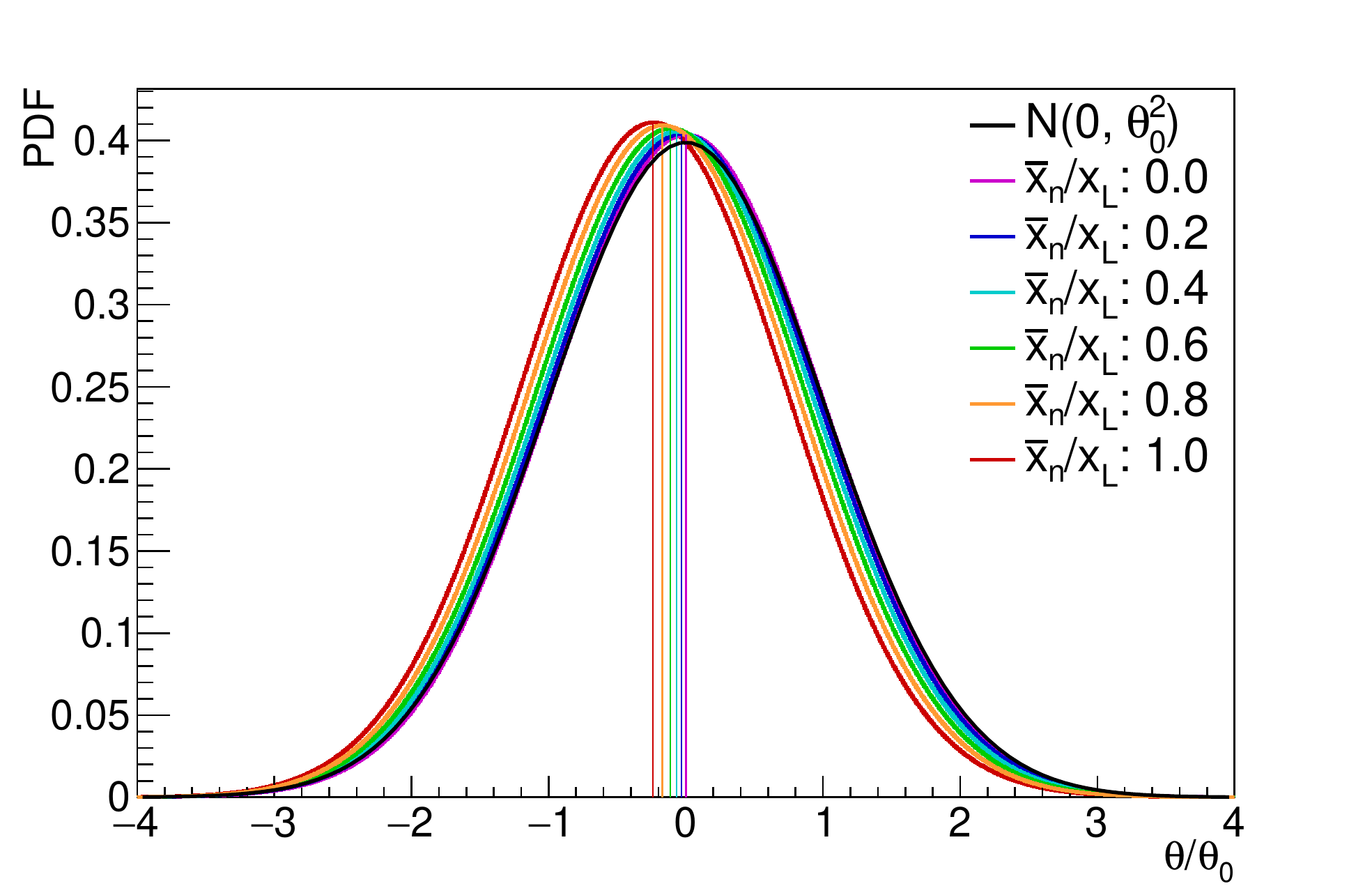}
		\caption{Deformation of the total sampled scattering angle distribution (black) for different true position mean, $\overline{x_n}$, at the n$^{\mathrm{th}}$ station.}
		\label{fig:angle}
	\end{minipage}
\end{figure}

The function in equation~\ref{eq:fxn} is represented for different values of $x_n/x_L$ in figure~\ref{fig:angle_xn}. At large $x_n$, the function, $f_{x_n}(\theta)$, is highly asymmetric due to scraping out of the fiducial volume. The choice of parameters is consistent with a simulated 280\,MeV/$c$ beam travelling between the two trackers at Step IV. The effective RMS scattering angle is $\theta_0\simeq9$\,mrad and the effective distance is $\Delta z/x_L\simeq17.5$.

The average over the fiducial distribution of $x_n$, $g(x_n)$, reads
\begin{eqnarray}
f(\theta) & = & C\int_{-\infty}^{+\infty}\frac{f_{x_n}(\theta)}{C_n}g(x_n)\mathrm{d}x_n \nonumber \\
& = & C\int_{-\infty}^{+\infty}\mathcal{N}(0,\theta_0^2)\bm{1}_{[-x_L,x_L]}\left(x_n+\Delta z\theta\right)\frac{1}{\sqrt{2\pi}\sigma_{x_n}}\exp\left[-\frac{1}{2}\frac{(x_n-\overline{x_n})^2}{\sigma_{x_n}^2}\right]\mathrm{d}x_n \nonumber \\
& = &\mathcal{N}(0,\theta_0^2)\times C\int_{-x_L-\theta\Delta z}^{x_L-\theta\Delta z}\frac{1}{\sqrt{2\pi}\sigma_{x_n}}\exp\left[-\frac{1}{2}\frac{(x_n-\overline{x_n})^2}{\sigma_{x_n}^2}\right]\mathrm{d}x_n \nonumber \\
& = & \mathcal{N}(0,\theta_0^2)\times\frac C2\left(\erf\left[\frac{x_L-\theta\Delta z-\overline{x_n}}{\sqrt{2}\sigma_{x_n}}\right]+\erf\left[\frac{x_L+\theta\Delta z+\overline{x_n}}{\sqrt{2}\sigma_{x_n}}\right]\right),
\label{eq:angle}
\end{eqnarray}
with $C^{-1}=\int_{-\infty}^{+\infty}f(\theta)\mathrm{d}\theta$. This shows a deviation from the non-biased scattering distribution $\mathcal{N}(0,\theta_0^2)$. In the limit $x_L/\sigma_{x_n}\rightarrow+\infty$, the normal distribution is recovered.

The function in equation~\ref{eq:angle} is represented for different values of $\overline{x_n}/x_L$ in figure~\ref{fig:angle}. For positive means, $\overline{x_n}$, the scattering angle distribution, $f(\theta)$, is biased towards negative values due to anisotropic scraping. At $\overline{x_n}=0$, the distribution is sharper than the normal $\mathcal{N}(0,\theta_0^2)$ and has tamed tails but retains the same mean. The large values of scattering angles are under-represented in the distribution due to inevitable transmission losses between measurement stations. The choice of parameters is motivated by the same requirements as for figure~\ref{fig:angle_xn}. The width of the distribution of $x_n$ is chosen to be $\sigma_{x_n}/x_L=0.5$.

At different limits, the formula behaves as expected:
\begin{equation}
\begin{array}{l}
 \lim\limits_{\overline{x_n}\rightarrow 0}f(\theta) = f(\theta)|_{\overline{x_n}=0} = f(-\theta)|_{\overline{x_n}=0}\,\forall\theta\rightarrow  \langle f(\theta)\rangle = \int_{0}^{+\infty}\theta \left[f(\theta)-f(-\theta)\right]\mathrm{d}\theta=0, \\
 \lim\limits_{x_L\rightarrow +\infty}f(\theta) = \mathcal{N}(0,\theta_0^2) \rightarrow \langle f(\theta)\rangle = 0, \\
 \lim\limits_{\sigma_{x_n}\rightarrow 0}f(\theta) = f_{\overline{x_n}}(\theta)\rightarrow\langle f(\theta)\rangle = -\sqrt{\frac{2}{\pi}}\theta_0\exp\left[-\frac{x_L^2+\overline{x_n^2}}{2\theta_0^2\Delta z^2}\right]\sinh\left[\frac{x_L\overline{x_n}}{\theta_0^2\Delta z^2}\right].
 \end{array}
 \label{eq:limits}
\end{equation}
A perfectly centred beam with $\overline{x_n}=0$ produces an even distribution in $\theta$. A detector large enough to contain the entire beam also yields an unbiased scattering angle distribution. A pencil beam still produces an angular distribution sensitive to the extrapolated position at the n$^\mathrm{th}$ sampling station as $\theta_0>0$.

\begin{figure}[!htb]
	\centering
	\includegraphics[width=.9\textwidth]{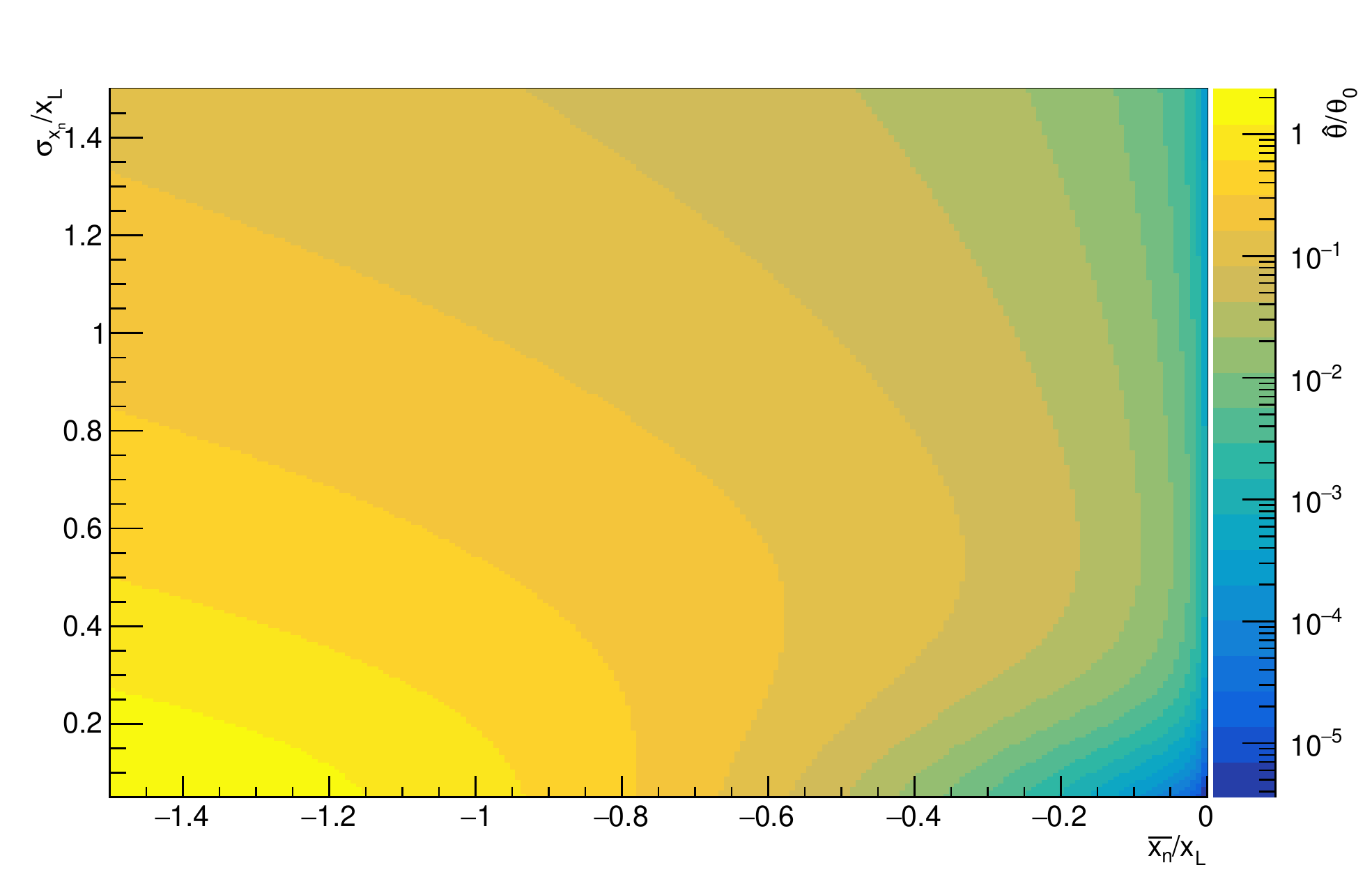}
	\caption{Sample mean of the scattering distribution with respect to the relative mean and width of the true position distribution at n$^{\mathrm{th}}$ station.}
	\label{fig:mean_angle_2d}
\end{figure}

Figure~\ref{fig:mean_angle_2d} represents the sample mean scattering angle, $\hat{\theta}$, normalised by the true effective scattering angle, $\theta_0$, as a function of the true position mean, $\overline{x_n}$, and width, $\sigma_{x_n}$. The figure is represented for negative values of $\overline{x_n}$ to produce log-friendly positive values of $\hat{\theta}/\theta_0$.

Provided that $\hat{\theta}\neq0$ for most beam settings, equation~\ref{eq:true_mean} in terms of sample means reads
\begin{equation}
    \langle x_n-\delta x_n-x_{n0}\rangle = \langle x_n-x_{n0}\rangle - \delta x_n = \langle f(\theta)\rangle\Delta z -\delta_n = \hat{\theta}\Delta z- \delta x_n. \\
\end{equation}
For common values of $\hat{\theta}>0.01\theta_0$ and $\theta_0=0.009$, the offset is of order $0.1$\,mrad, which is larger than a tenth of the expected Tait-Bryan angles of the tracker modules. This constitutes a significant bias on the measurement of the station offset, $\delta_n$, and must be tackled to ensure a consistent measurement of the offset independent of the MICE beam setting. 

\subsection{Criterion}
\label{sec:criterion}
The beam anisotropy is the main source of bias, as demonstrated in section~\ref{sec:sampling}. An isotropic particle sample is selected out to ensure a consistent measurement of the alignment. 

It is not possible to make a selection depending on the track position at TOF1 and TOF2. A sample that makes it to TOF2 is inevitably biased by requiring that the particles made it to TOF2. In field-off data, 80\% of the particles scrape out before reaching TOF2. TOF1 does not provide measurements of the track gradients and as such cannot be used standalone.

The upstream tracker is chosen to make a selection as it is the most upstream detector that measures the position and gradients of the track. The criterion is based on the likelihood of the particle to scatter out of the fiducial volume before it makes it to the centre of the downstream tracker. This analysis requests a 90\,\% probability of containment, i.e. $\sim2.15\,\sigma$ for a 2D Gaussian. For an effective distance $\Delta z$ between the two trackers, the fiducial radius $R_L$ and an RMS effective scattering angle $\theta_0$, the criterion is expressed as
\begin{equation}
(\xi_u^d-x_D)^2+(\upsilon_u^d-y_D)^2 < (R_L-2.15\theta_0\Delta_z)^2,
\label{eq:criterion}
\end{equation}
with $(\xi_u^d,\upsilon_u^d)$ the global upstream track propagated to TKD and $(x_D,y_D)$ the downstream tracker centre. This is equivalent to selecting out a smaller circular fiducial surface at the centre of TKD. It ensures containment upstream and downstream so that $\sigma_{x_n}/x_L\ll1$, $\overline{x_n}/x_L\ll1$ and $\hat{\theta}\simeq0$ in equation\,\ref{eq:limits}.

\begin{figure}[!htb]
	\begin{minipage}[b]{.45\textwidth}
		\centering
		\begin{tikzpicture}[scale=.8]
		\draw[thick,->] (4, 0) -- (-4, 0) node[anchor=north east] {x};
		\draw[thick,->] (0, -4) -- (0, 4) node[anchor=south west] {y};
		
		\draw[thick] (-0.1, 3) -- (0.1, 3) node[anchor=south west] {$R_L$};
		\draw[thick] (-0.1, 1.8) -- (0.1, 1.8) node[anchor=south west] {$R_L-2.15\theta_0\Delta z$};
		
		\draw (0,0) circle(3);
		\draw[fill=blue, draw=black, fill opacity = 0.2] (0,0) circle (1.8);
		
		\draw[fill=black, draw=black] (-2, -2) circle (0.05) node[anchor=north west]{$(\xi_1^u, \upsilon_1^u)$};
		\draw[dashed, ->] (-2, -2) -- (-1.1, -1.1);
		\draw[fill=black, draw=black] (-1, -1) circle (0.05) node[anchor=south west]{$(\xi_1^d, \upsilon_1^d)$};
		
		\draw[fill=black, draw=black] (1, -0.5) circle (0.05) node[anchor=north west]{$(\xi_2^u, \upsilon_2^u)$};
		\draw[dashed, ->] (1, -0.5) -- (2.4, 0.4);
		\draw[fill=black, draw=black] (2.5, 0.5) circle (0.05) node[anchor=south west]{$(\xi_2^d, \upsilon_2^d)$};
		\end{tikzpicture}
	\end{minipage}
	\hfill
	\begin{minipage}[b]{.45\textwidth}
		\centering
		\begin{tikzpicture}[scale=.8]
		\draw[thick,->] (4, 0) -- (-4, 0) node[anchor=north east] {x};
		\draw[thick,->] (0, -4) -- (0, 4) node[anchor=south west] {y};
		
		\draw[thick] (-0.1, 3) -- (0.1, 3) node[anchor=south west] {$R_L$};
		
		\draw (0,0) circle(3);
		\draw[fill=green, draw=black, fill opacity = 0.2] (-1,-1) circle (1.2);
		\draw[fill=red, draw=black, fill opacity = 0.2] (2.5,0.5) circle (1.2);
		
		\draw[fill=black, draw=black] (-1, -1) circle (0.05) node[anchor=south west]{$(\xi_1^d, \upsilon_1^d)$};
		\draw[fill=black, draw=black] (2.5, 0.5) circle (0.05) node[anchor=south west]{$(\xi_2^d, \upsilon_2^d)$};
		\end{tikzpicture}
	\end{minipage}
	\caption{\textbf{Left}: Schematic of the total (white) and selected (blue) tracker fiducial surface in the $xy$ plane. The movement of two example particles are represented. \textbf{Right}: $2.15\,\sigma$ position ellipses for two example particles.}
	\label{fig:cont}
\end{figure}
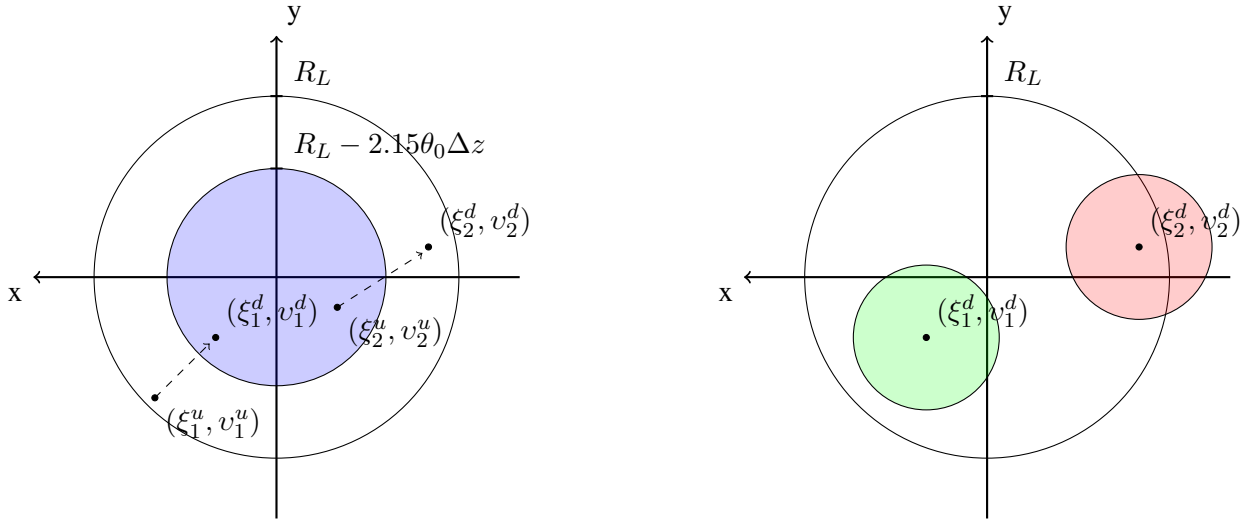

Figure~\ref{fig:cont} shows a geometrical representation of the selection criterion. The phase space of two particles, $P_1$ and $P_2$, are sampled at the upstream tracker and their parameters ($\xi_i^u, \upsilon_i^u, \xi_i^{'u}, \upsilon_i^{'u}$) are extracted. Their expected position in the downstream tracker are extrapolated as $\psi_i^d=\psi_i^u+\psi_i^{'u}(\zeta_d-\zeta_u)$, $\psi=\xi,\upsilon$. If the downstream point belongs to the blue fiducial surface on the left-hand drawing, as for $P_1$, the particle is kept in the final sample. The right-hand figure represents the $2.15\,\sigma$ ellipses of the particles' expected position due to scattering between the two trackers. The ellipse is contained for the accepted particle, $P_1$, but is not for the rejected particle, $P_2$. The inclusion of the green ellipse ensures at least a 90\% probability that the particle goes through the downstream tracker.

The RMS effective scattering angle, $\theta_0$, and effective distance, $\Delta z$, in equation~\ref{eq:criterion} are functions of the momentum, $p$, the particle species and the amount of material, $dz/X_0$. The practicalities of evaluating these quantities and applying the criterion will be described in section~\ref{sec:impcrit} for a Monte Carlo sample.

\subsection{Outliers}
\label{sec:outliers}
The criterion limits the number of particles that scrape between TKU and TKD as it selects a tight aperture in the downstream tracker. Due to the large non-Gaussian tails of the scattering distribution and decays in flight, a non-negligible fraction of particles get lost in transmission. A beam anisotropic about the downstream tracker center scrapes preferentially on one end of the tracker, introducing a bias. The lost particles must be accounted for in the residual distributions.

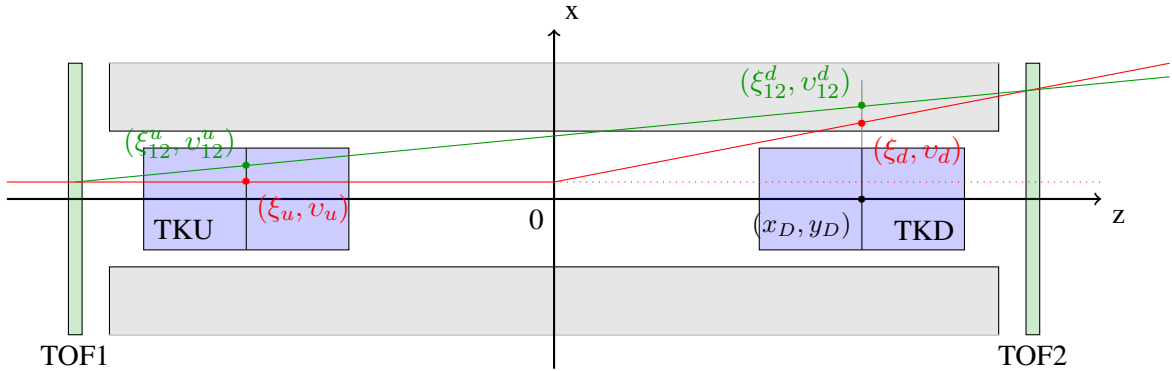
\begin{figure}[!htb]
	\centering
	\begin{tikzpicture}[scale=0.9]
	\draw[fill=black, fill opacity=0.1] (-6.5, 1) rectangle (6.5, 2);	
	\draw[white] (-6.5, 2) -- (6.5, 2);
	\draw[fill=black, fill opacity=0.1] (-6.5, -1) rectangle (6.5, -2);
	\draw[white] (-6.5, -2) -- (6.5, -2);	
	
	\draw[thick,->] (-8, 0) -- (8, 0) node[anchor=north west] {z};
	\draw[thick,->] (0, -2.5) -- (0, 2.5) node[anchor=south west] {x};
	\draw (0, 0) node[below left] {0};
	
	\draw[black, fill=green!60!black, fill opacity=0.2] (-7.1, -2) rectangle (-6.9, 2);	
	\draw (-7, -2) node [below] {TOF1};
	\draw[black, fill=green!60!black, fill opacity=0.2] (6.9, -2) rectangle (7.1, 2);	
	\draw (7, -2) node [below] {TOF2};
	\draw[black, fill=blue, fill opacity=0.2] (-6, -0.75) rectangle (-3, 0.75);
	\draw (-6, -0.75) node [above right] {TKU};
	\draw (-4.5, -0.75) -- (-4.5, 0.75);
	\draw[black, fill=blue, fill opacity=0.2] (3, -0.75) rectangle (6, 0.75);
	\draw (6, -0.75) node [above left] {TKD};
	\draw (4.5, -0.75) -- (4.5, 0.75);
	\draw (4.5, -0.05) node {\small\textbullet};
	\draw (4.5, -0.05) node[below left] {\small$(x_D,y_D)$};
	\draw[gray] (4.5, 0.75) -- (4.5, 1.75);
	
	\draw[red] (-4.5, 0.22) node[centered] {\small\textbullet};
	\draw[red] (-4.5, 0.22) node[below right] {$(\xi_u,\upsilon_u)$};
	
	\draw[green!60!black] (-4.5, 0.46) node[centered] {\small\textbullet};
	\draw[green!60!black] (-4.5, 0.46) node[above left] {$(\xi_{12}^u,\upsilon_{12}^u)$};	
	
	\draw[red] (4.5, 1.075) node[centered] {\small\textbullet};
	\draw[red] (4.5, 1.075) node[below right] {$(\xi_d,\upsilon_d)$};
	
	\draw[green!60!black] (4.5, 1.335) node[centered] {\small\textbullet};
	\draw[green!60!black] (4.5, 1.335) node[above left] {$(\xi_{12}^d,\upsilon_{12}^d)$};
	
	\draw[red] (-8, 0.25) -- (0, 0.25);
	\draw[red] (0, 0.25) -- (9, 2);
	\draw[red, dotted] (0, 0.25) -- (8, 0.25);
	
	\draw[green!60!black] (-7, 0.25) -- (9, 1.8);
	
	\end{tikzpicture}
	\caption{Paraxial track (red line) that undergoes a hard scatter at the level of the absorber ($z=0$) and hits the magnet bore. The track interpolated between the two time-of-flight stations is represented by the green line.}
	\label{fig:scrape}	
\end{figure}

Figure\,\ref{fig:scrape} shows a paraxial particle that is expected to be transmitted but has a hard scatter at the level of the absorber ($z=0$). In this scenario, no track is recorded in the downstream tracker. If the extrapolated transversal position in TKD, $\psi_u^d$, $\psi=\xi,\upsilon$, is farther on one side of the tracker centre, it is more likely to scatter out on the same side, as it requires a smaller, more probable scattering angle. To account for those particles, outliers are added to the residual distributions with signs chosen with the following prescriptions
\begin{equation}
	\psi_u^d > q_D\rightarrow \\
	\left\{
	\begin{array}{l}
	\psi_u-\psi_{12}^u < 0 \\
	\psi'_u-\psi_{12}' < 0 \\
	\psi_d-\psi_{12}^d < 0 \\
	\psi_d-\psi_{12}' > 0
	\end{array} \right.,\qquad
	\psi_u^d < q_D\rightarrow \\
	\left\{
	\begin{array}{l}
	\langle\psi_u-\psi_{12}^u\rangle > 0 \\
	\langle\psi'_u-\psi_{12}'\rangle > 0 \\
	\langle\psi_d-\psi_{12}^d\rangle > 0 \\
	\langle\psi_d-\psi_{12}'\rangle < 0
	\end{array}
	\right.,\qquad\psi=\xi,\upsilon,\,q=x,y.
\end{equation} 

\section{Alignment of a Monte Carlo sample}
\label{sec:single_beam}
Two main beam line settings are used in order to produce the alignment corrections. High momentum beams are preferred in order to reduce the RMS scattering angle and maximize transmission. The Monte Carlo reproduces two runs taken during the ISIS 2017/01 user cycle presented in table\,\ref{tab:settings}. The two settings correspond to `pion' beams of positive polarity to maximize statistics. The number of particle triggers has been greatly enhanced in the Monte Carlo in order to investigate the resolution of the alignment algorithm.

\begin{table}[!h]
\centering
\begin{tabular}{c|c|c|c|c|c|c}
Run ID & Date & Beam type & $\langle p_u\rangle$ [MeV/c] & $\sigma_{p_u}$ [MeV/c] & TOF1 SPs & MC TOF1 SPs \\
\hline
9367 & 30/05/17 & $\pi^+$ & 285.00$\pm$0.03 & 11.53$\pm$0.02 & 186136 & 2056605 \\
9370 & 30/05/17 & $\pi^+$ & 394.40$\pm$0.03 & 13.55$\pm$0.02 & 133253 & 767107
\end{tabular}
\caption{Characteristics of runs 9367 and 9370. The mean and RMS upstream momentum are reconstructed for the MC from the velocity of the muons between TOF0 and TOF1. SP stands for space point.}
\label{tab:settings}
\end{table}

The 9367 Monte Carlo sample is separated out into ten chunks and the 9370 into five. The chunks are combined together at the calibration stage but are fitted separately during the alignment stage in order to verify the consistency of the algorithm.

The two beams exhibit different distributions in the beam line. The mean position of the former lies in the negative $x$ and negative $y$ while the latter lies in the positive quadrant. Any bias that could arise from the beam characteristics is amplified by the use of these two settings. An agreement between the two independent fits guarantees an unbiased measurement of the alignment constants.

\subsection{Trackers internal alignment}
As a consistency check, the alignment of the tracker stations with respect to each other is measured. As for the CMM measurements described in section\,\ref{sec:surveys}, the two end stations (1 and 5) are used as the reference for the alignment and the positions of the other stations are measured with respect to them.

\begin{figure}[!htb]
	\begin{minipage}[b]{.49\textwidth}
		\centering
		\includegraphics[width=\textwidth]{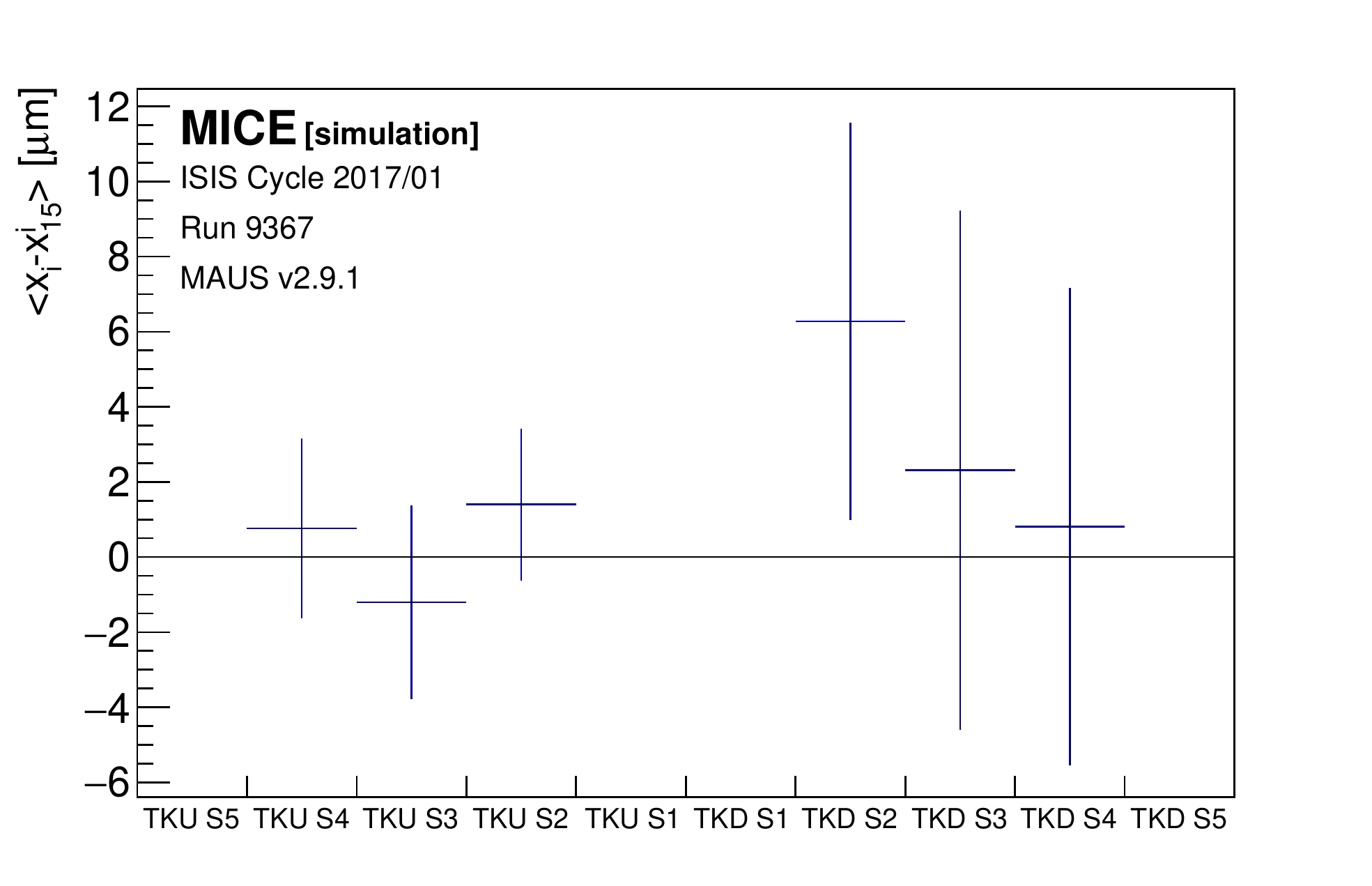}
	\end{minipage}
	\hfill
	\begin{minipage}[b]{.49\textwidth}
		\centering
		\includegraphics[width=\textwidth]{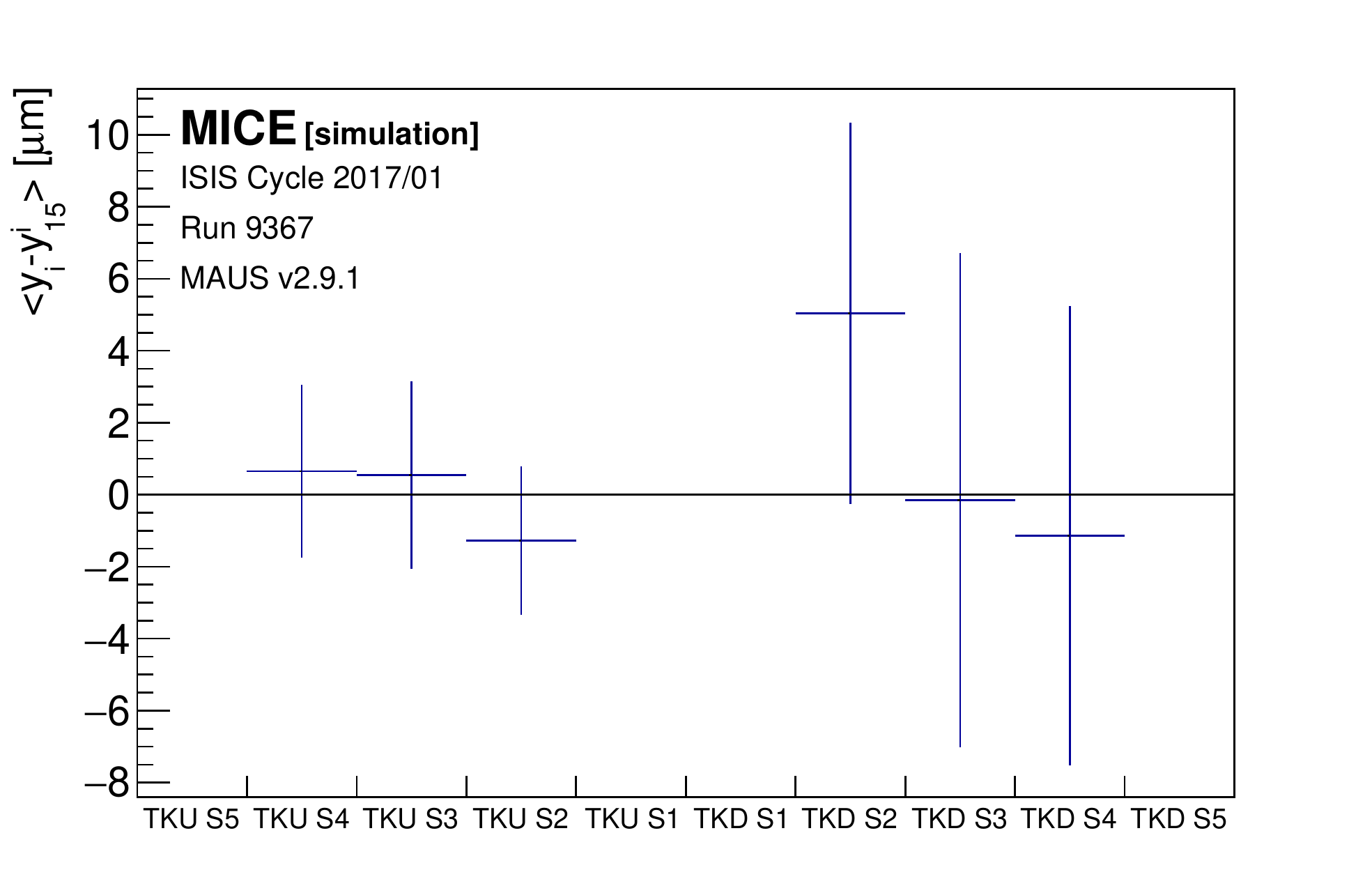}
	\end{minipage}
		\caption{Mean residuals between the interpolated track joining the two outermost stations of each tracker and the position reconstructed in the central three stations in $x$ (\textbf{left}) and $y$ (\textbf{right}).}
		\label{fig:stat}	
\end{figure}

For each track recorded in tracker $t=u,d$, the reference axis gradients are reconstructed as $q'_{15}=(q_5-q_1)/(z_5-z_1)$, $q=x,y$, and the $y$-intercept is the position recorded in station 1. The interpolated position in station $i$ reads
\begin{equation}
q_{15}^i=q_1+q'_{15}(z_i-z_1).
\end{equation}
The mean residuals $\langle q_{15}^i-q_i\rangle$, $q=x,y$, are represented in figure\,\ref{fig:stat} from the most upstream to the most downstream station. As expected from aligned stations, the mean residuals are all compatible with zero with a resolution of order 1\,$\mu$m. This technique can be used to align the tracker stations using real data down to a precision two orders of magnitude smaller than the tracker space point resolution.

\subsection{Implementation of the criterion}
\label{sec:impcrit}
Applying the criterion in equation~\ref{eq:criterion} requires the determination of the mean RMS effective scattering angle for a given particle, $\theta_0^i$. It is function of the upstream particle momentum, $p_u$, its velocity, $\beta_u$, and the amount of material traversed between the two trackers, $dz/X_0$.

\begin{figure}[!htb]
	\begin{minipage}[b]{.475\textwidth}
		\centering
		\includegraphics[width=\textwidth]{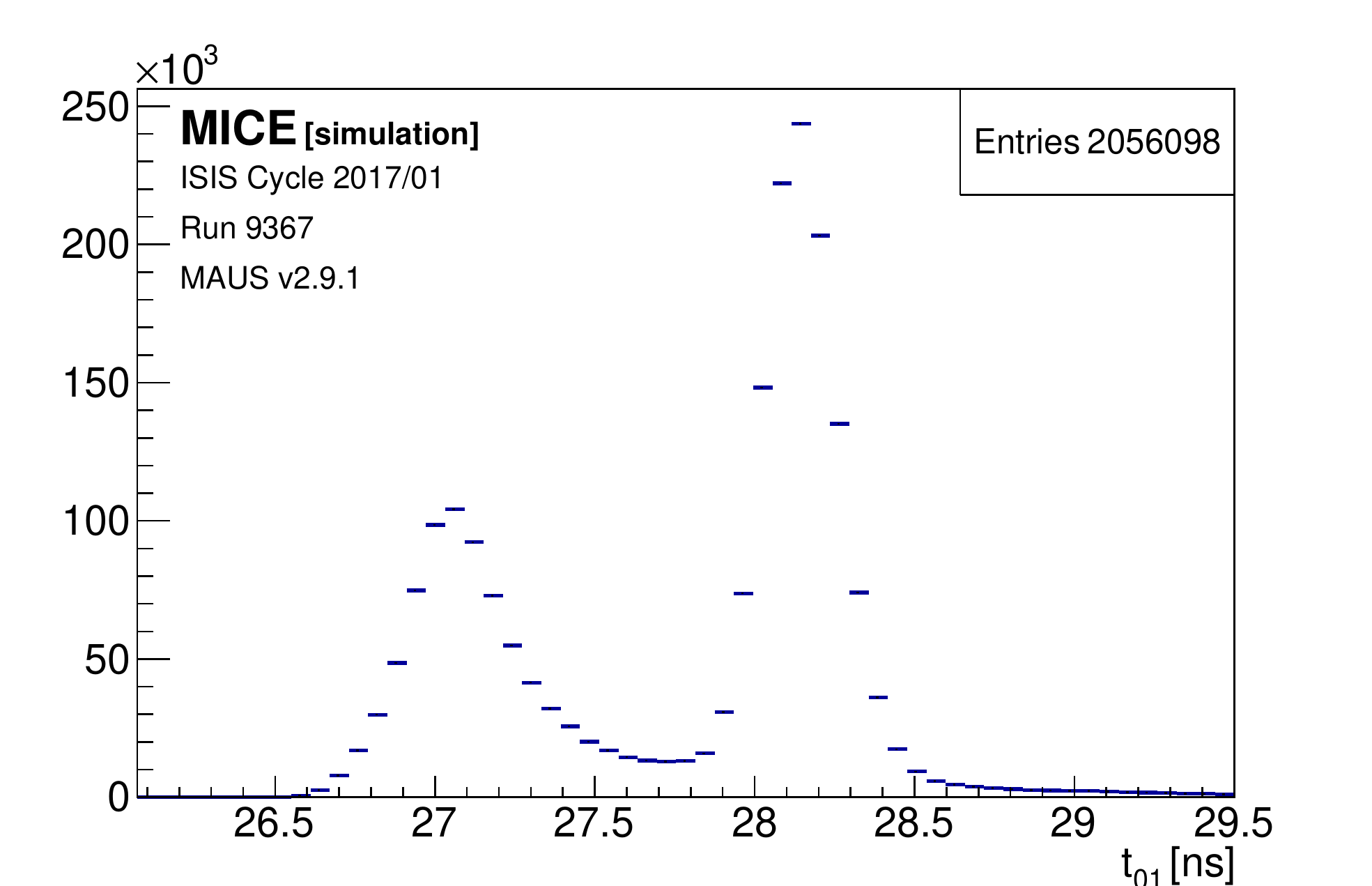}
		\caption{Time-of-flight distribution of a pion beam between TOF0 and TOF1.\vspace{0.575cm}}
		\label{fig:tof01}
	\end{minipage}
	\hfill
	\begin{minipage}[b]{.475\textwidth}
		\centering
		\includegraphics[width=\textwidth]{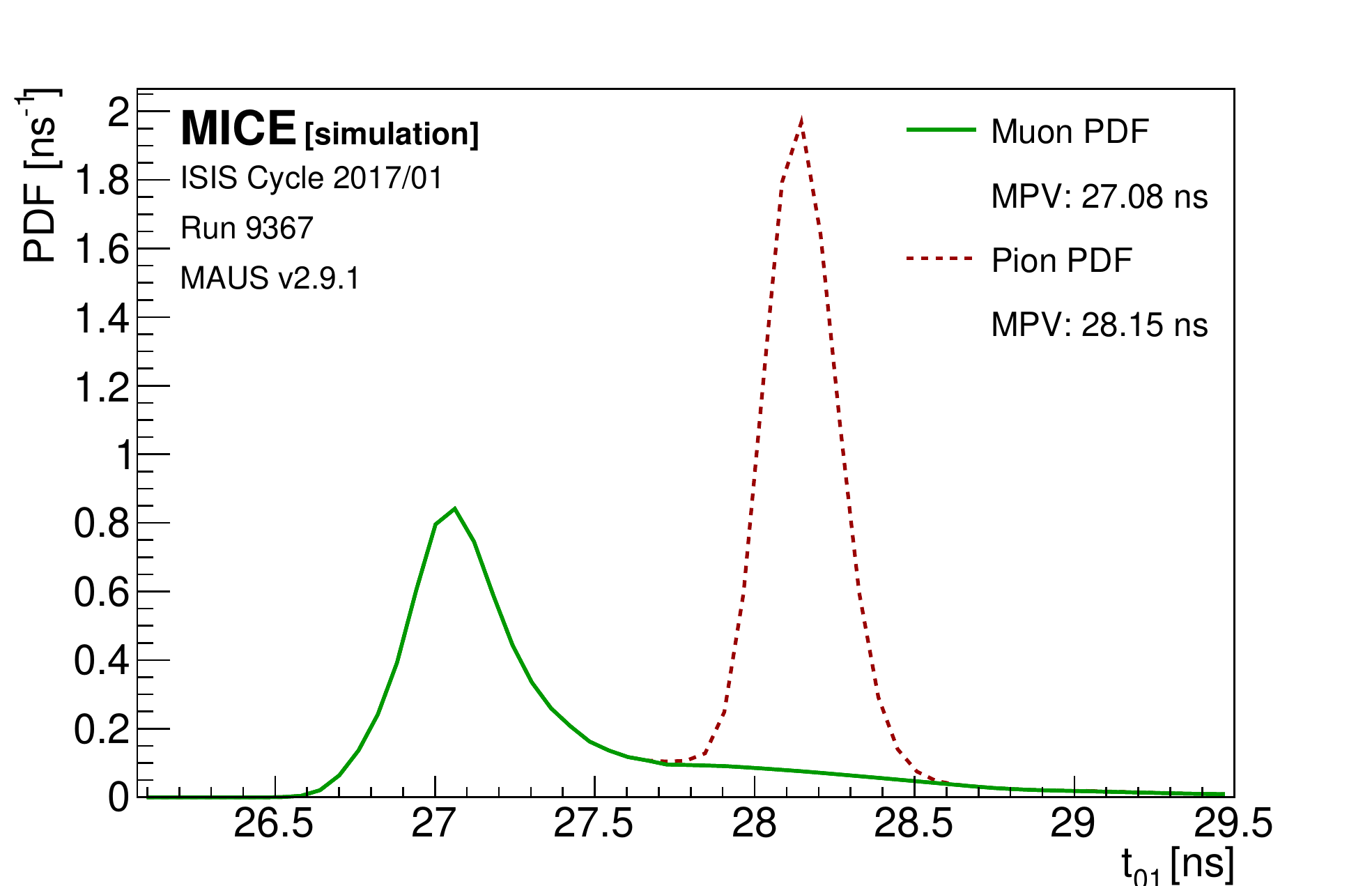}
		\caption{Normalised probability density functions of the muon and pion components of the beam. MPV stands for most probable value.}
		\label{fig:tof01_pid}
	\end{minipage}
\end{figure}

The time-of-flight between TOF0 and TOF1, $t_{01}$, provides a strong particle identification variable. In the pion beams used for the alignment, three species of particles ($\pi^+,\mu^+,e^+$) are present in the sample but have distinguishable time-of-flight distributions as shown in figure~\ref{fig:tof01}. Positrons are rejected from this analysis as they are ultrarelativistic and their momentum cannot be reconstructed from the time-of-flight. Requiring $t_{01}>26$\,ns rids the sample of positrons.

The remainder of the distribution contains three components. The bulk of muons and pions are contained in the left and right peaks, respectively. The muons produced by pion decays in flight form a decreasing time-of-flight continuum that reaches a probability maximum at the level of the muon peak. The decay muons are discriminated by using background identification methods~\cite{tspectrum_bg}. They are subtracted from the distribution and the rest of it is fitted with a two-peak Gaussian to extract the locations, $t_\mu$ and $t_\pi$, and widths, $\sigma_\mu$ and $\sigma_\pi$, of the muon and pion peaks, respectively. The background-subtracted peaks do not overlap. The intermediary time-of-flight, $t_c$, of equal probability for the two peaks is located at 
\begin{equation}
\frac{t_c-t_\mu}{\sqrt{2}\sigma_\mu} = \frac{t_\pi-t_c}{\sqrt{2}\sigma_\pi} \iff t_c = \frac{\sigma_\mu t_\pi +\sigma_\pi t_\mu}{\sigma_\mu+\sigma_\pi}.
\end{equation}
The pion PDF lies above $t_c$. The background is added to the remainder of the distribution to form the muon PDF. The two normalised distributions are represented stacked on top of each other in figure\,\ref{fig:tof01_pid}. For a given test sample, the particle species is selected to be the most probable species.

Given the species $i$, the momentum is inferred from the time-of-flight through
\begin{equation}
p_{01}=\frac{m_ic}{\sqrt{\left(ct_{01}/s_{01}\right)^2-1}},
\end{equation}
with $s_{01}$ the distance between the space points in TOF0 and TOF1. The momentum is adjusted for the loss in TOF1 by subtracting the Bethe-Bloch mean energy loss going through the scintillating material. This yields the upstream momentum, $p_u$. The upstream velocity is derived from the momentum through
\begin{equation}
\beta_u = \frac{p_u}{\sqrt{p_u^2+m_i^2c^2}}.
\end{equation}

\begin{figure}[!htb]
\begin{minipage}[b]{.49\textwidth}
\centering
\includegraphics[width=\textwidth]{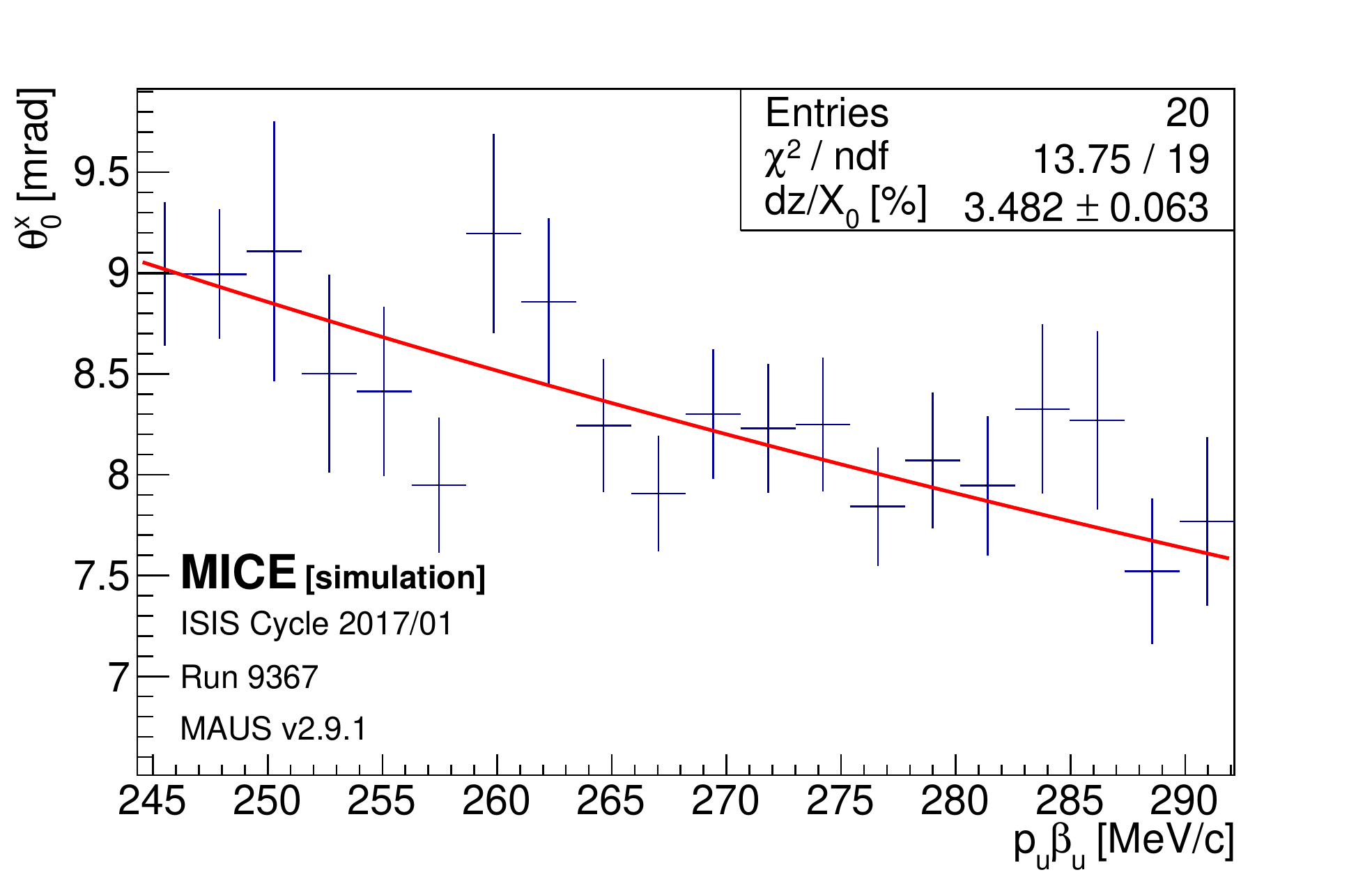}
\end{minipage}
\hfill
\begin{minipage}[b]{.49\textwidth}
\centering
\includegraphics[width=\textwidth]{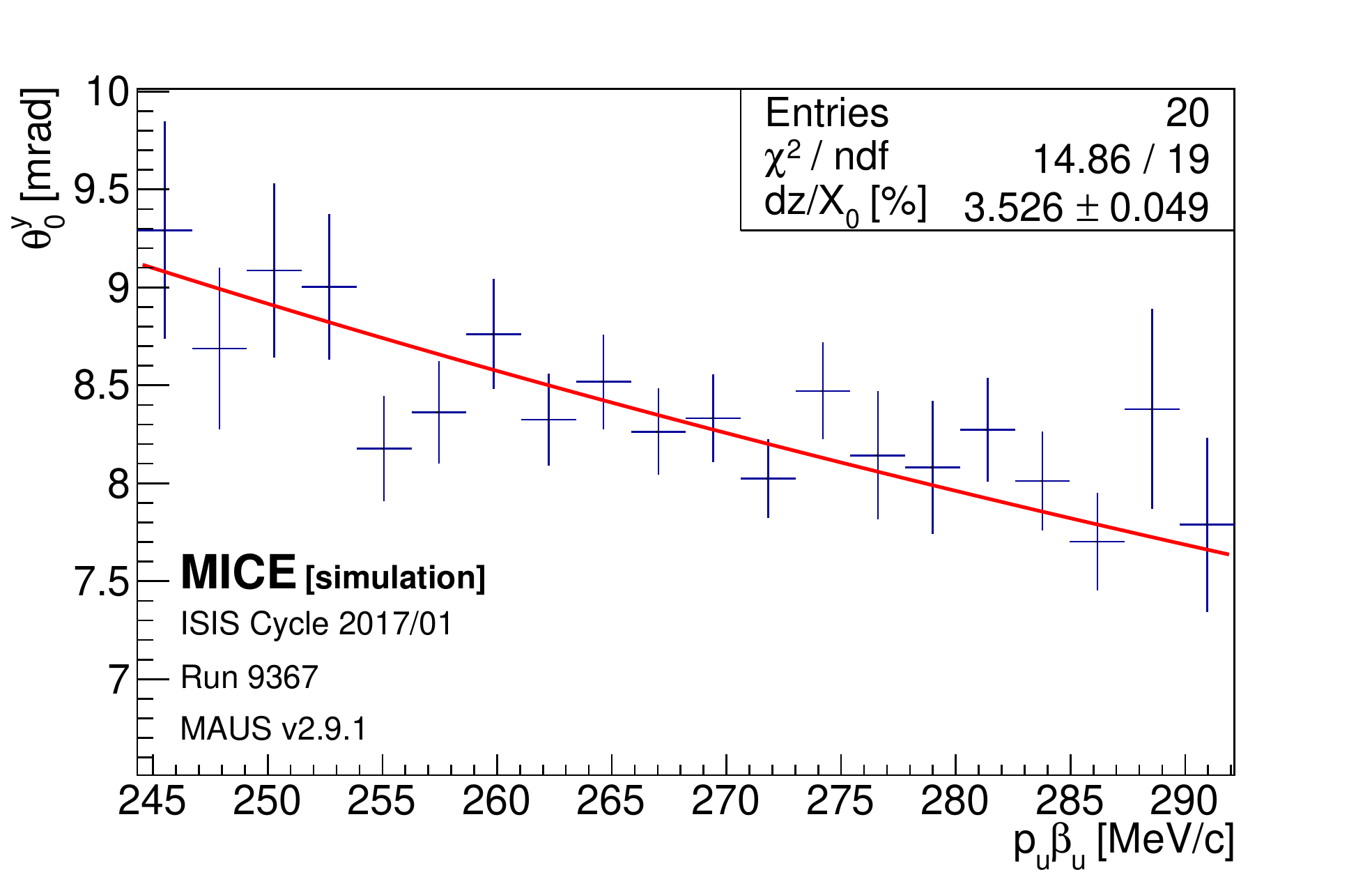}
\end{minipage}
\caption{Theoretical fit to the RMS scattering angle in the $xz$ and $yz$ projections between the upstream and downstream trackers as a function of the momentum and velocity reconstructed at the upstream tracker.}
\label{fig:dzx0}
\end{figure}

The value of path length in amount of radiation length, $dz/X_0$, is obtained by fitting the theoretical scattering width in equation~\ref{eq:criterion} to real data. Figure~\ref{fig:dzx0} shows the value of the RMS scattering angle, $\theta_0$ as a function of the momentum, $p_{12}$, in the $xz$ and $yz$ projections. These fits yield a value of the path length of $dz/X_0=3.509 \pm 0.039$\,\%, which produces a simplified expression for the RMS scattering angle:
\begin{equation}
\theta_0\simeq\frac{(2.223\,\text{MeV})}{\beta_{u} p_{u}c}.
\end{equation}

\begin{figure}[!htb]
\begin{minipage}[b]{.475\textwidth}
\centering
\includegraphics[width=\textwidth]{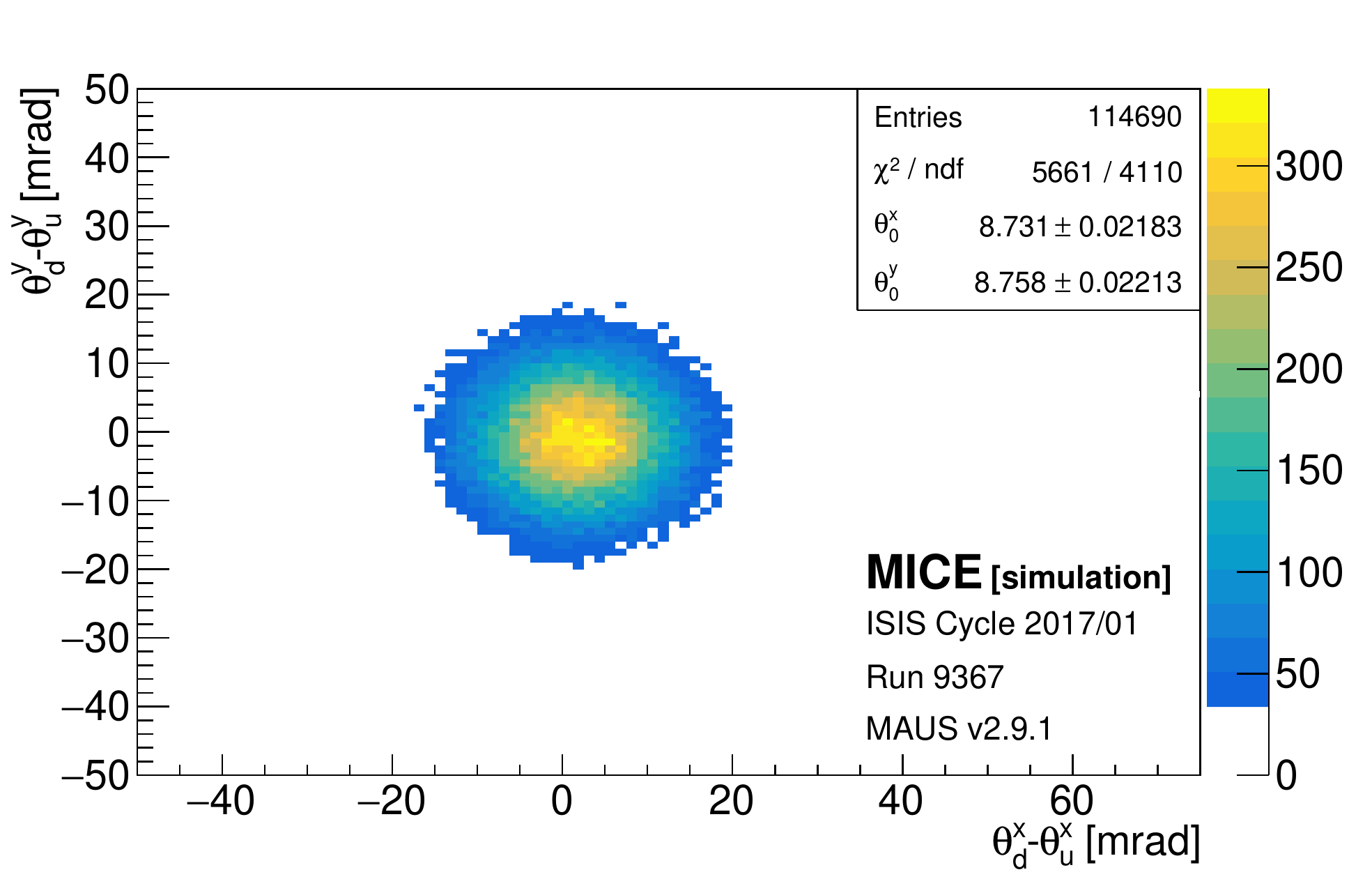}
\caption{Angular residuals between the gradients measured in TKD and TKU.}
\label{fig:resang}
\end{minipage}
\hfill
\begin{minipage}[b]{.475\textwidth}
\centering
\includegraphics[width=\textwidth]{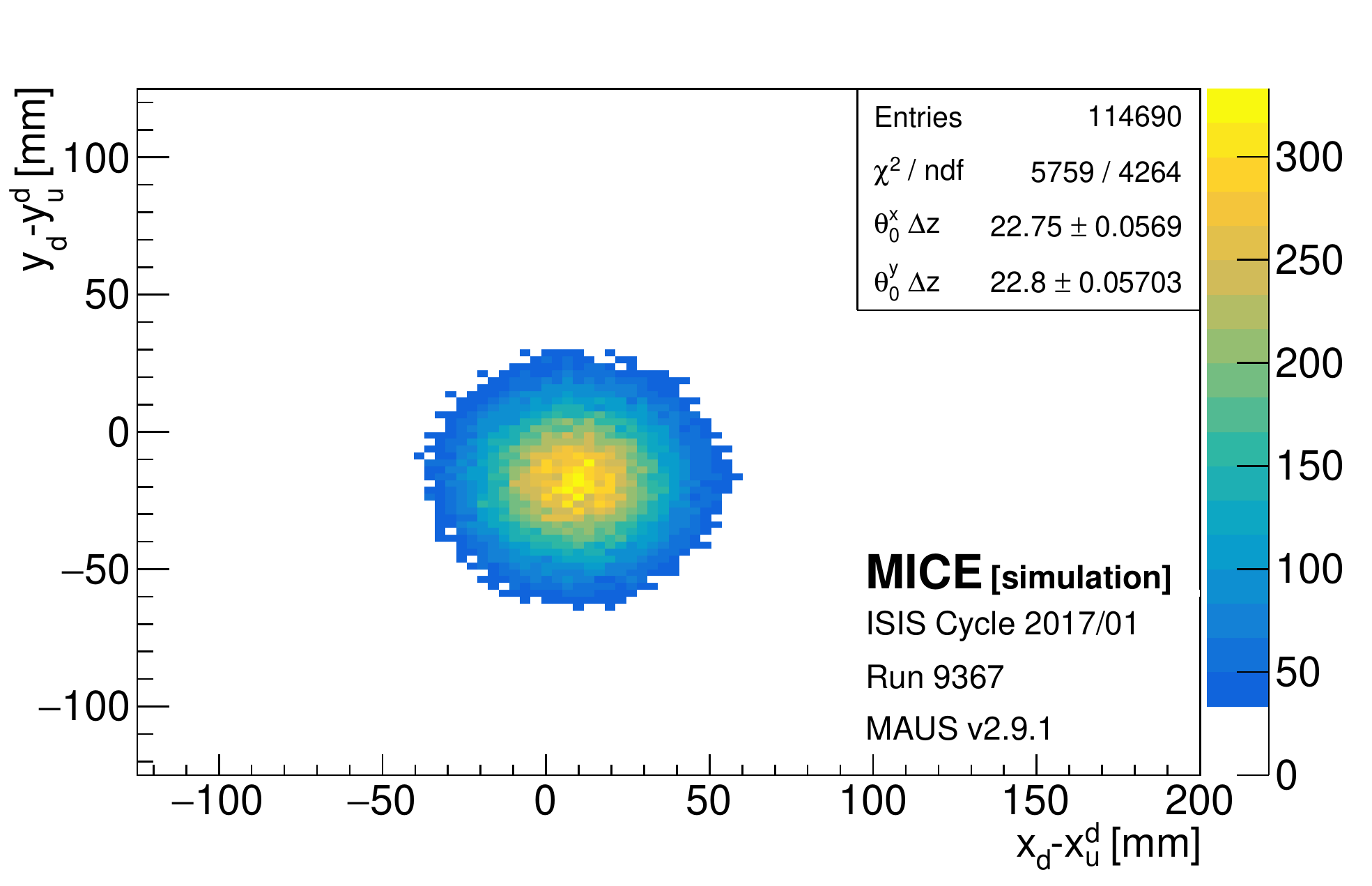}
\caption{Residuals between the position measured in TKD and the TKU position projected into TKD.}
\label{fig:respos}
\end{minipage}
\end{figure}

The last element to determine is the effective distance, $\Delta z$, i.e the constant that relates the effective scattering angle $\theta_0$ to the displacement in the downstream tracker, as defined in equation\,\ref{eq:effective}. To measure the RMS scattering angle, $\theta_0$, the angular residuals between the two trackers are measured as shown in figure~\ref{fig:resang}. This yields a weighted average of $\theta_0=8.74\pm0.02$\,mrad, consistent with this run setting mean momentum. The RMS of the residuals between the position measured in the centre of the downstream tracker and the location of the track extrapolated from the upstream tracker provides a measurement of $\theta_0\Delta z=22.77\pm0.04$\,mm, as shown in figure~\ref{fig:respos}. The effective distance between the two trackers is
\begin{equation}
\Delta z=\frac{\theta_0\Delta_z}{\theta_0}=2604.26\pm6.53\,\mathrm{mm},
\end{equation}
a few centimeters over half the physical distance between two tracker centres of 4886.57\,mm. Most of the scattering happens at the level of the absorber windows equidistant from both trackers. Provided with the set of equations and parameters determined in this section, the criterion is applied on a particle-by-particle basis to produce the analysis sample.

\begin{figure} [!h]
\centering
\begin{minipage}[b]{.49\textwidth}
\centering
\includegraphics[width=\textwidth]{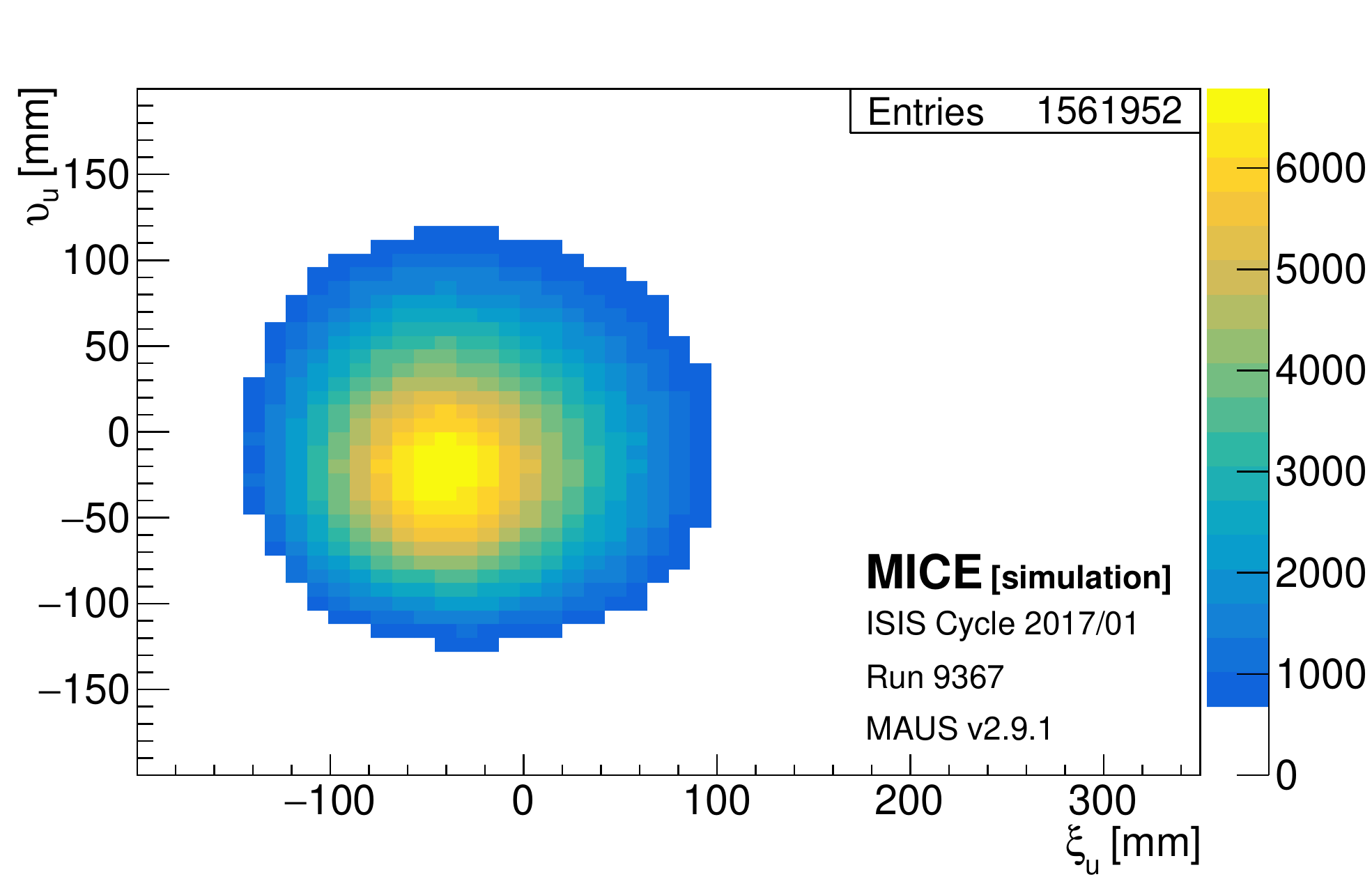}
\end{minipage}
\hfill
\begin{minipage}[b]{.49\textwidth}
\centering
\includegraphics[width=\textwidth]{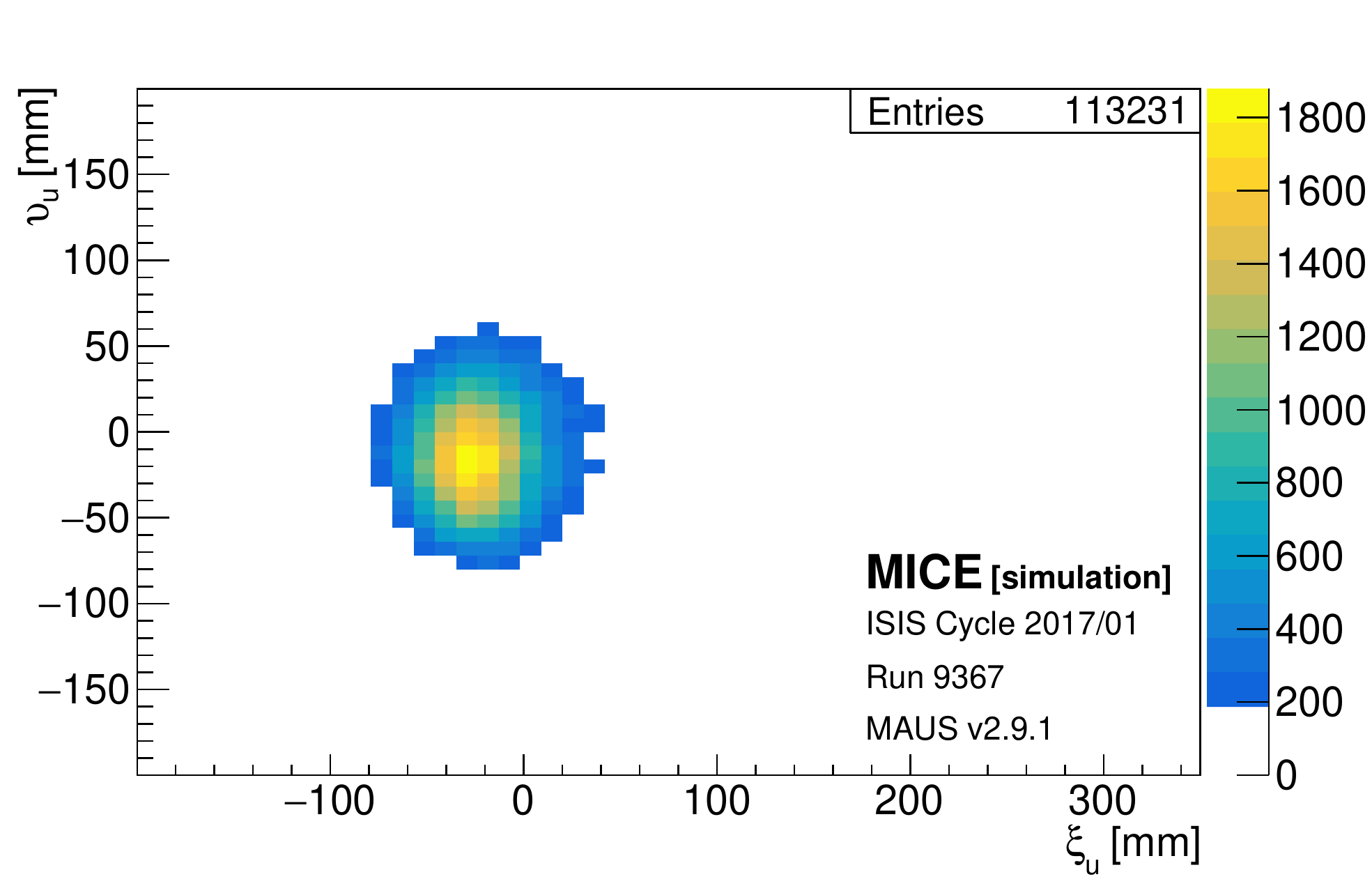}
\end{minipage}

\begin{minipage}[b]{.49\textwidth}
\centering
\includegraphics[width=\textwidth]{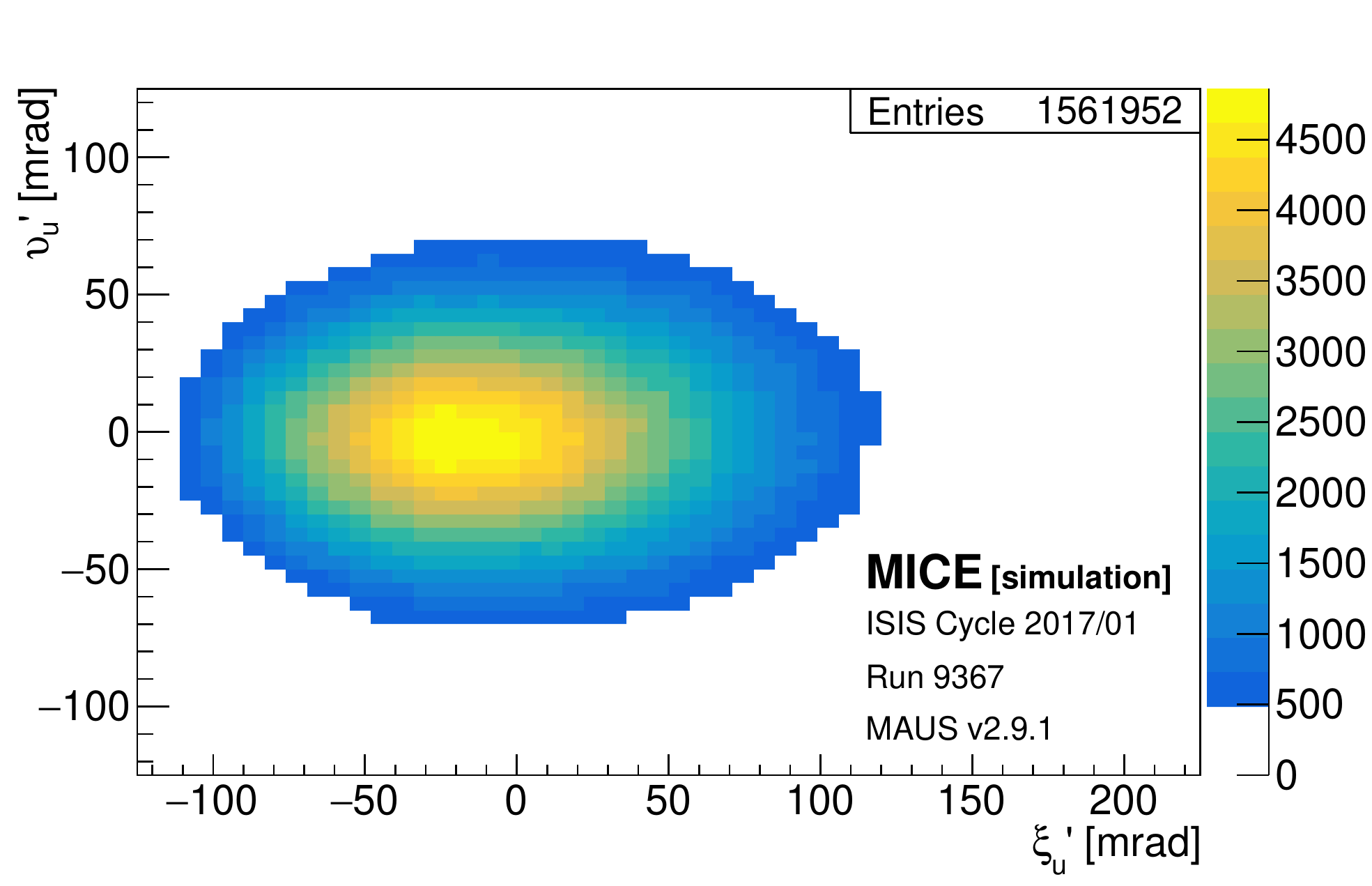}
\end{minipage}
\hfill
\begin{minipage}[b]{.49\textwidth}
\centering
\includegraphics[width=\textwidth]{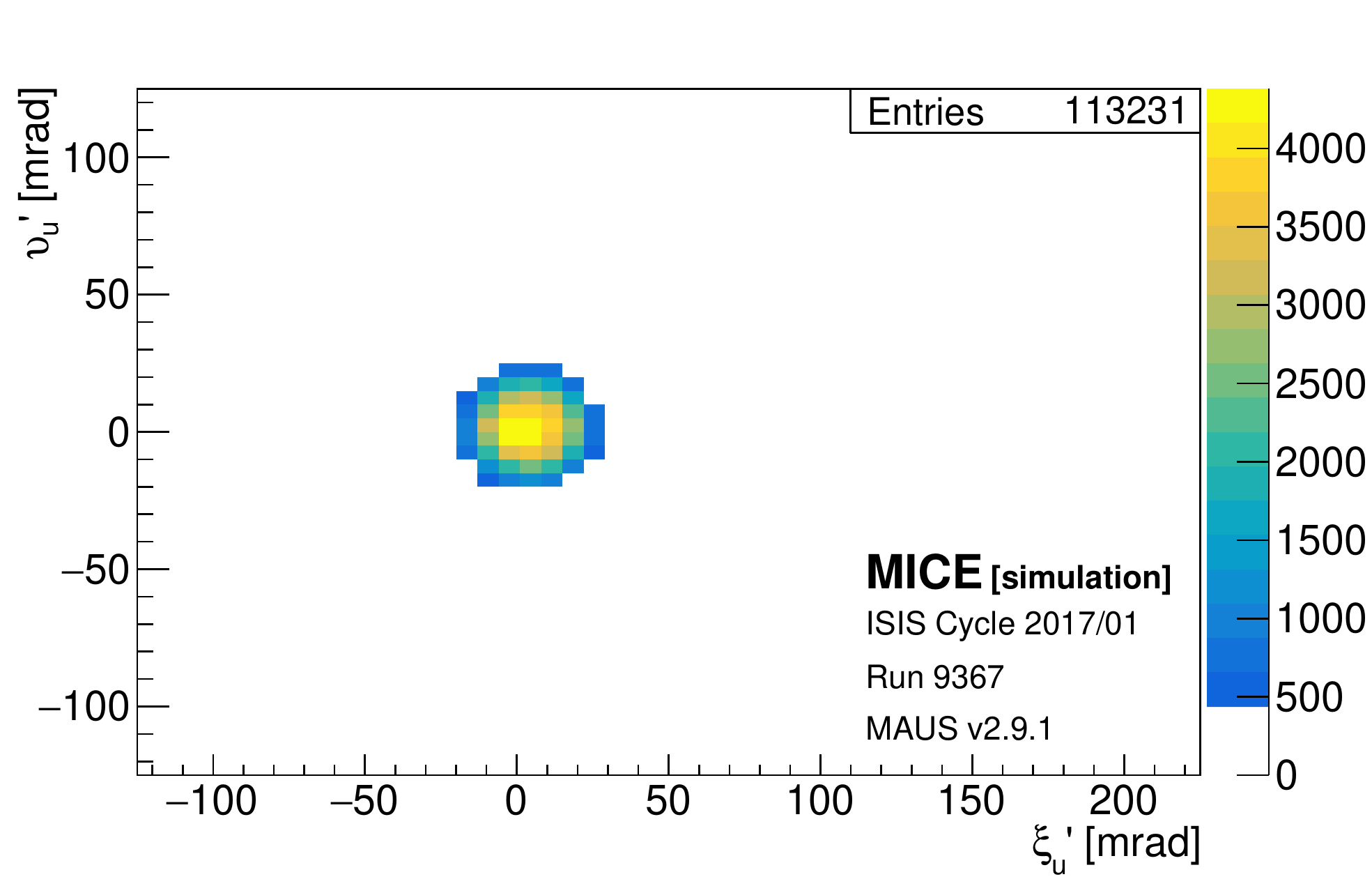}
\end{minipage}

\begin{minipage}[b]{.49\textwidth}
\centering
\includegraphics[width=\textwidth]{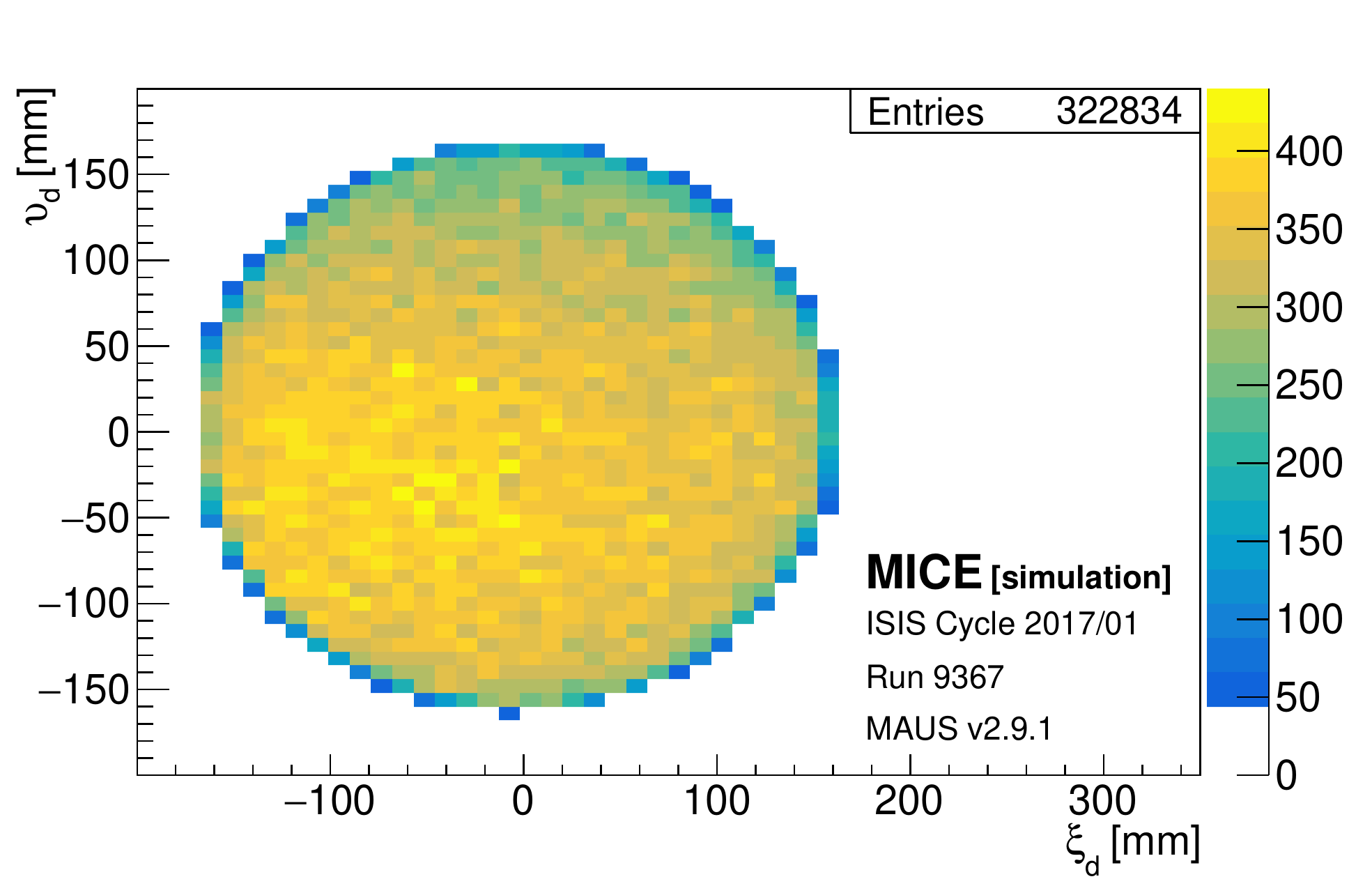}
\end{minipage}
\hfill
\begin{minipage}[b]{.49\textwidth}
\centering
\includegraphics[width=\textwidth]{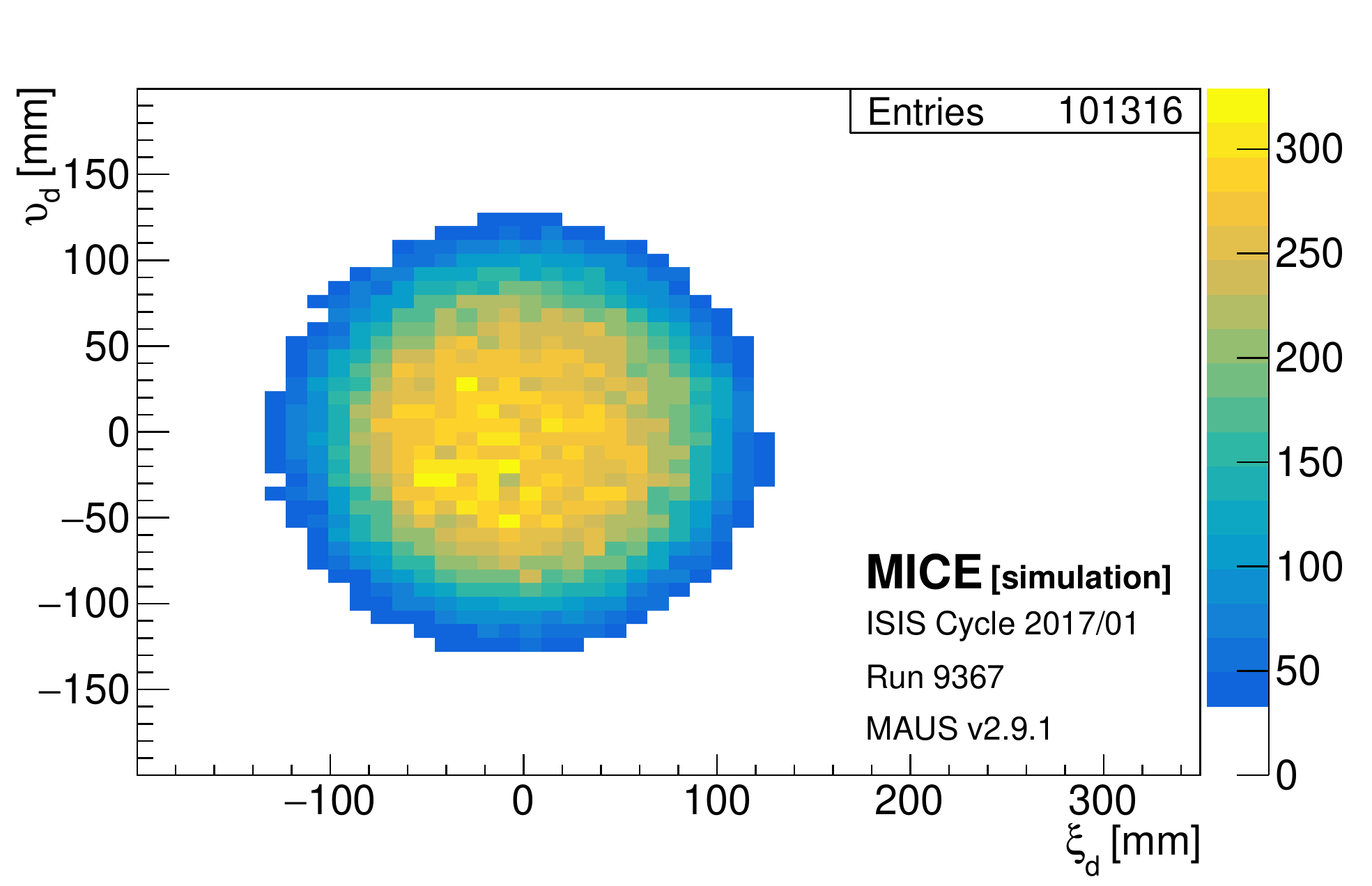}
\end{minipage}

\begin{minipage}[b]{.49\textwidth}
\centering
\includegraphics[width=\textwidth]{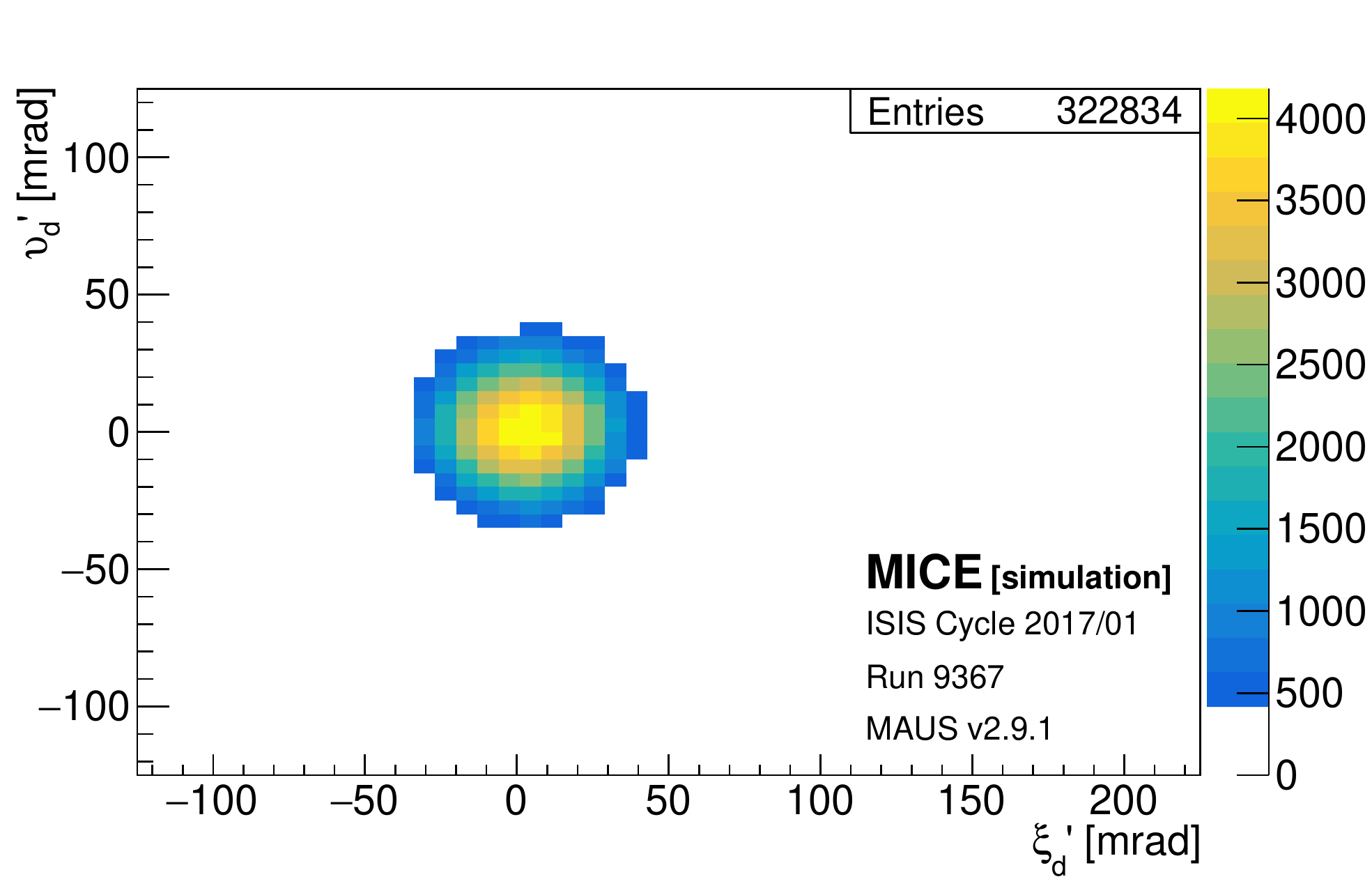}
\end{minipage}
\hfill
\begin{minipage}[b]{.49\textwidth}
\centering
\includegraphics[width=\textwidth]{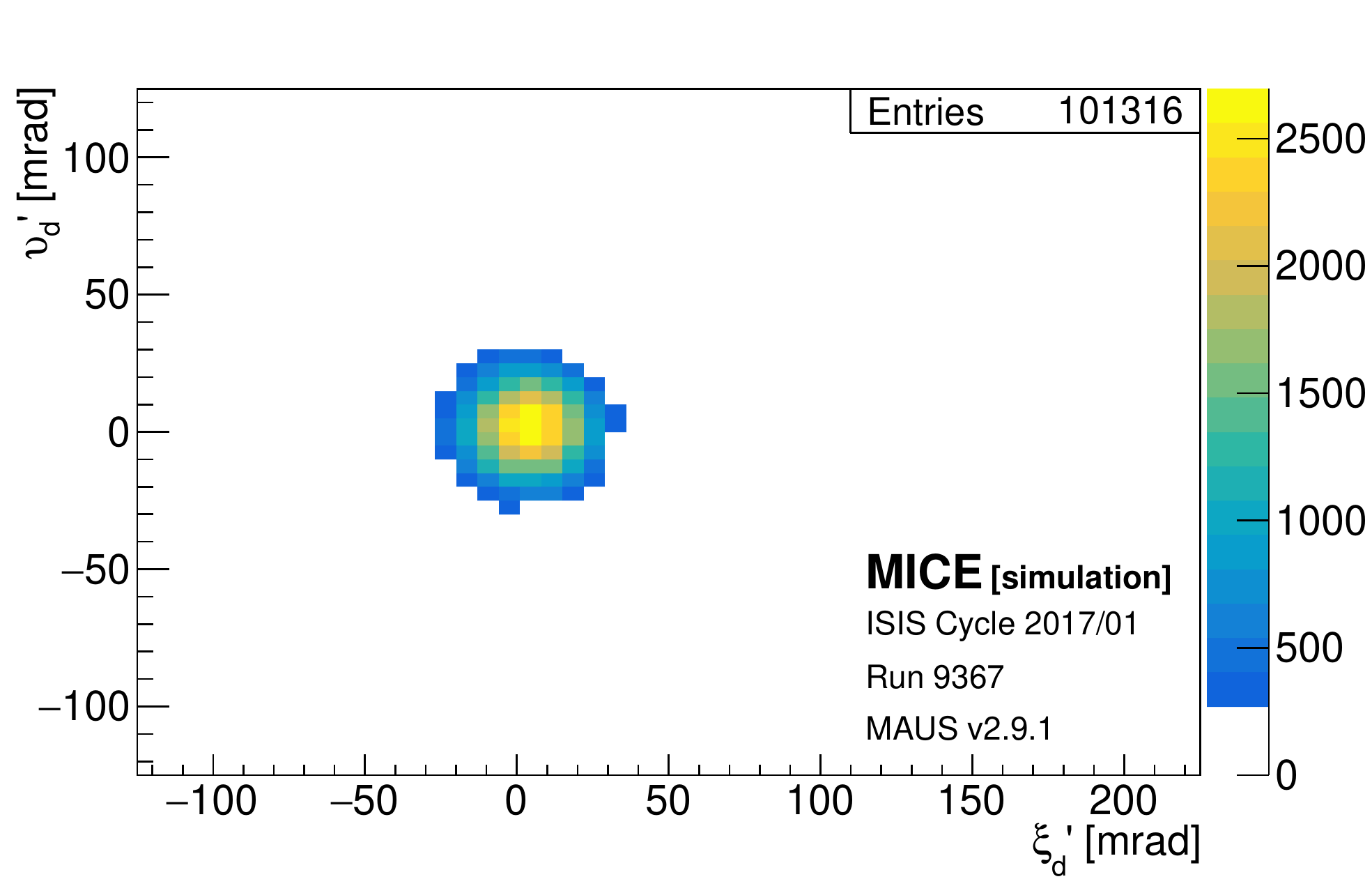}
\end{minipage}
\caption{Beam $(\xi_t,\upsilon_t)$ and $(\xi'_t,\upsilon'_t)$, $t=u,d$, profiles in the upstream (TKU) and downstream tracker (TKD) shown before (\textbf{left}) and after (\textbf{right}) the application of the selection criterion.}
\label{fig:profiles}
\end{figure}

The effect of this selection on the beam profile in the centre of the trackers is shown in figure~\ref{fig:profiles}. The criterion produces a very tight beam upstream that ensures that the sample is mostly contained downstream.

\subsection{Fitting algorithm}\label{sec:fitting_algo}
Provided with the unbiased sample produced as described in sections\,\ref{sec:criterion}--\ref{sec:outliers}, each track yields of a set of global gradients between TOF1 and TOF2, $\xi'_{12}$ and $\upsilon'_{12}$, and global extrapolated positions at the tracker centres, $\xi_{12}^t$ and $\upsilon_{12}^t$. It also records the position of the track at the centre of the trackers in local coordinates, $x_t$ and $y_t$, and its local gradients, $x'_t$ and $y'_t$. 

\begin{figure} [!htb]
	\begin{minipage}[b]{.48\textwidth}
		\centering
		\includegraphics[width=\textwidth]{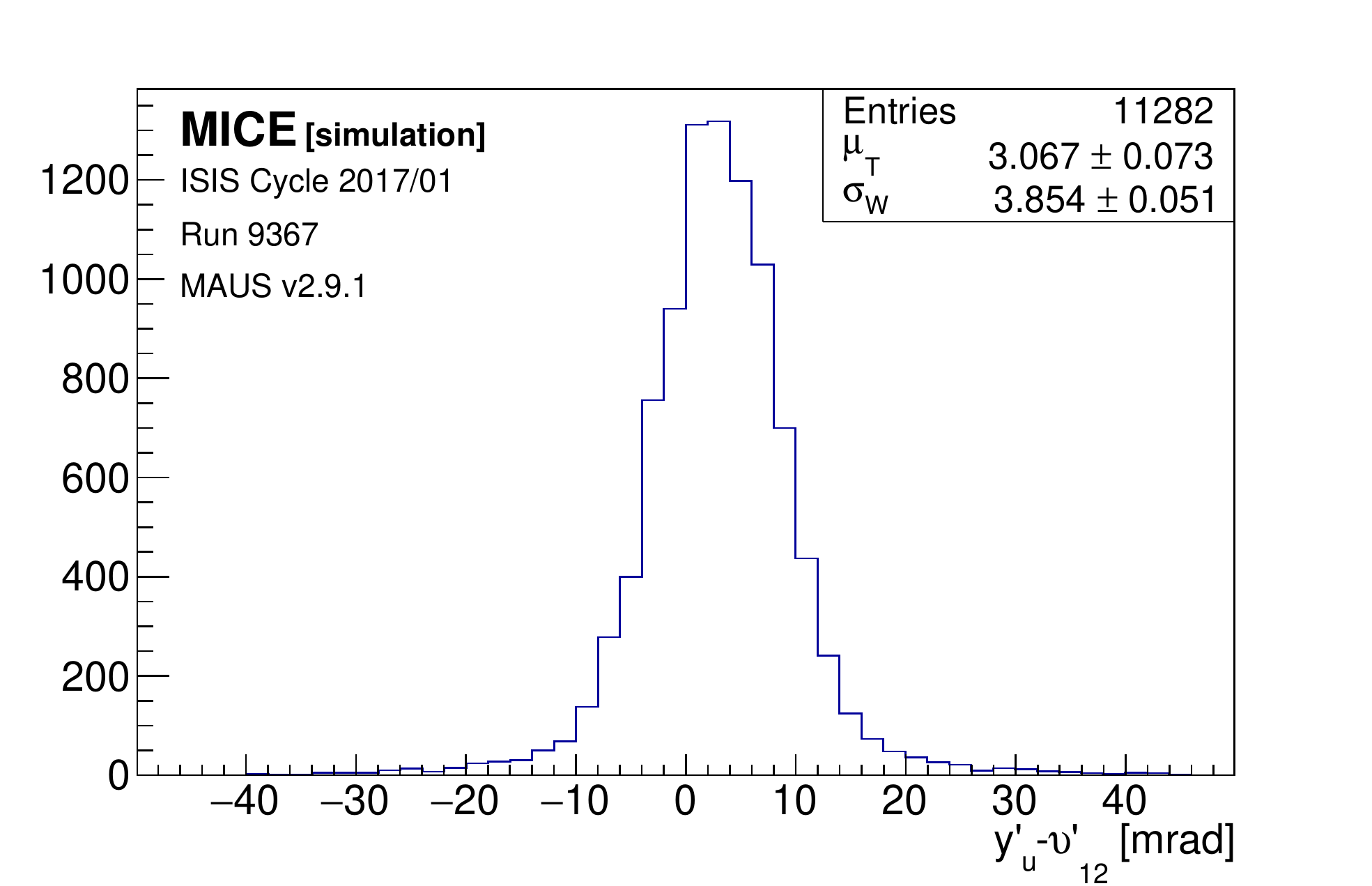}
		\caption{Residuals distribution between the pitch gradients measured locally in TKU, $y'_u$, and globally between TOF1 and TOF2, $\upsilon'_{12}$.}
		\label{fig:resyp}
	\end{minipage}
	\hfill
	\begin{minipage}[b]{.48\textwidth}
		\centering
		\includegraphics[width=\textwidth]{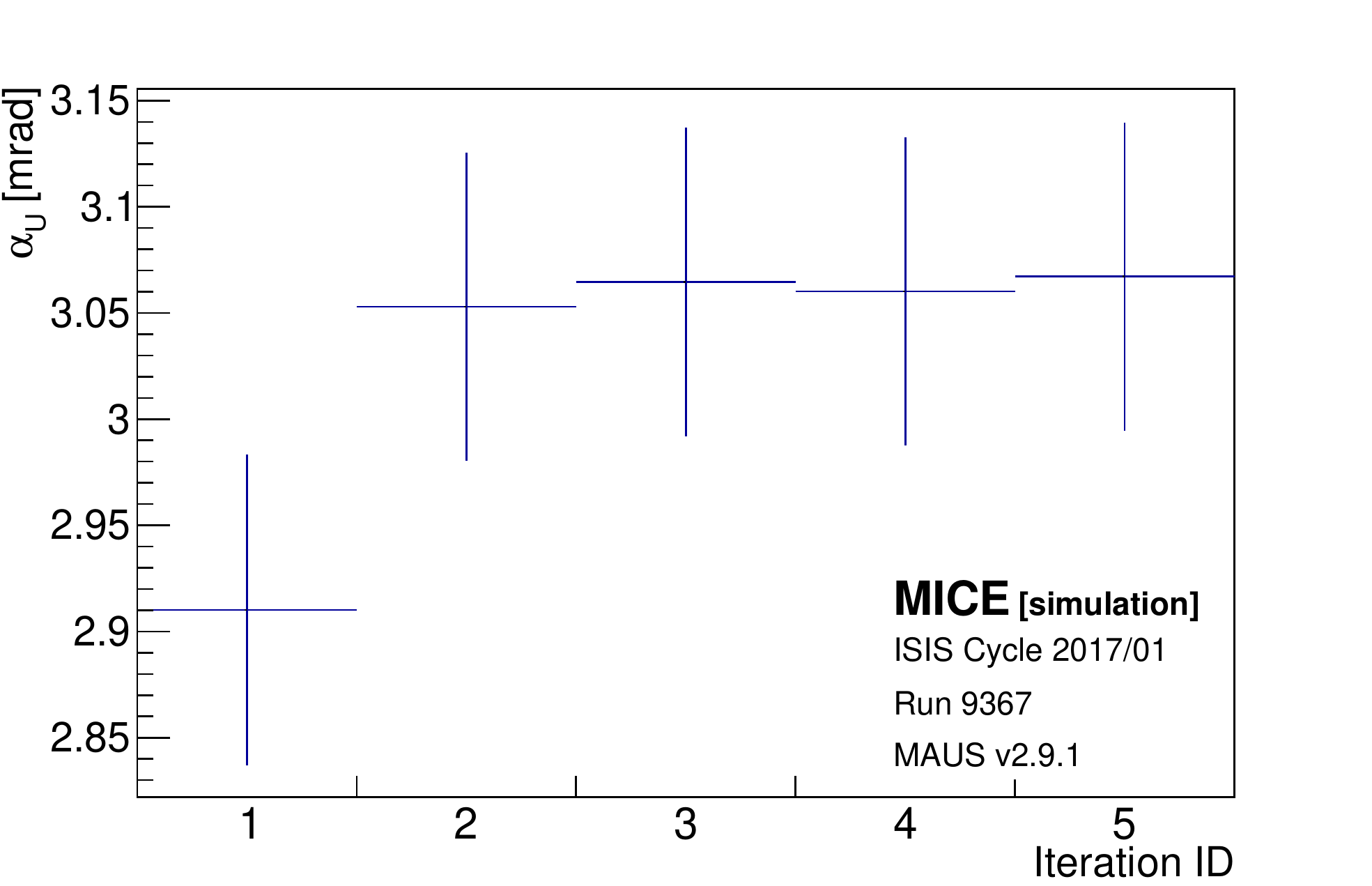}
		\caption{Evolution of the optimal value of the pitch angle in TKU, $\alpha_U^*$, for different number of iterations of the fitting algorithm.}
		\label{fig:optalpha}
	\end{minipage}
\end{figure}

The residual distributions necessary to measure the left hand side of equations\,\ref{eq:res} are produced in order to measure the eight alignment parameters. Figure~\ref{fig:resyp} shows the gradient residuals between $y'_u$ and $\upsilon'_{12}$. The mean residual yields the the pitch of the upstream tracker, $\alpha_U$. Particular care must be taken in the evaluation of the mean, as the scattering distribution exhibits long tails and the peak is not necessarily Gaussian as the the time-of-flight detectors have a very coarse granularity. The interquartile mean, more robust than the arithmetic mean, is used as figure-of-merit. The statistical uncertainty carried by the measurement is $2\sigma_W/\sqrt{N}$, with $\sigma_W$ the interquantile Winsorised RMS~\cite{winsor}.

To ensure the best possible fit to the tracker parameters, the algorithm is applied multiple times. The first estimate of $x_T,y_T,\alpha_T,\beta_T$ is used as an input to the sample selection part of the algorithm. The process is repeated until the alignment constants converge. Figure~\ref{fig:optalpha} shows the evolution of the optimal upstream tracker pitch, $\alpha_U^
*$, over five iterations.

\begin{figure} [!htb]
	\centering
	\begin{minipage}[b]{.49\textwidth}
		\centering
		\includegraphics[width=\textwidth]{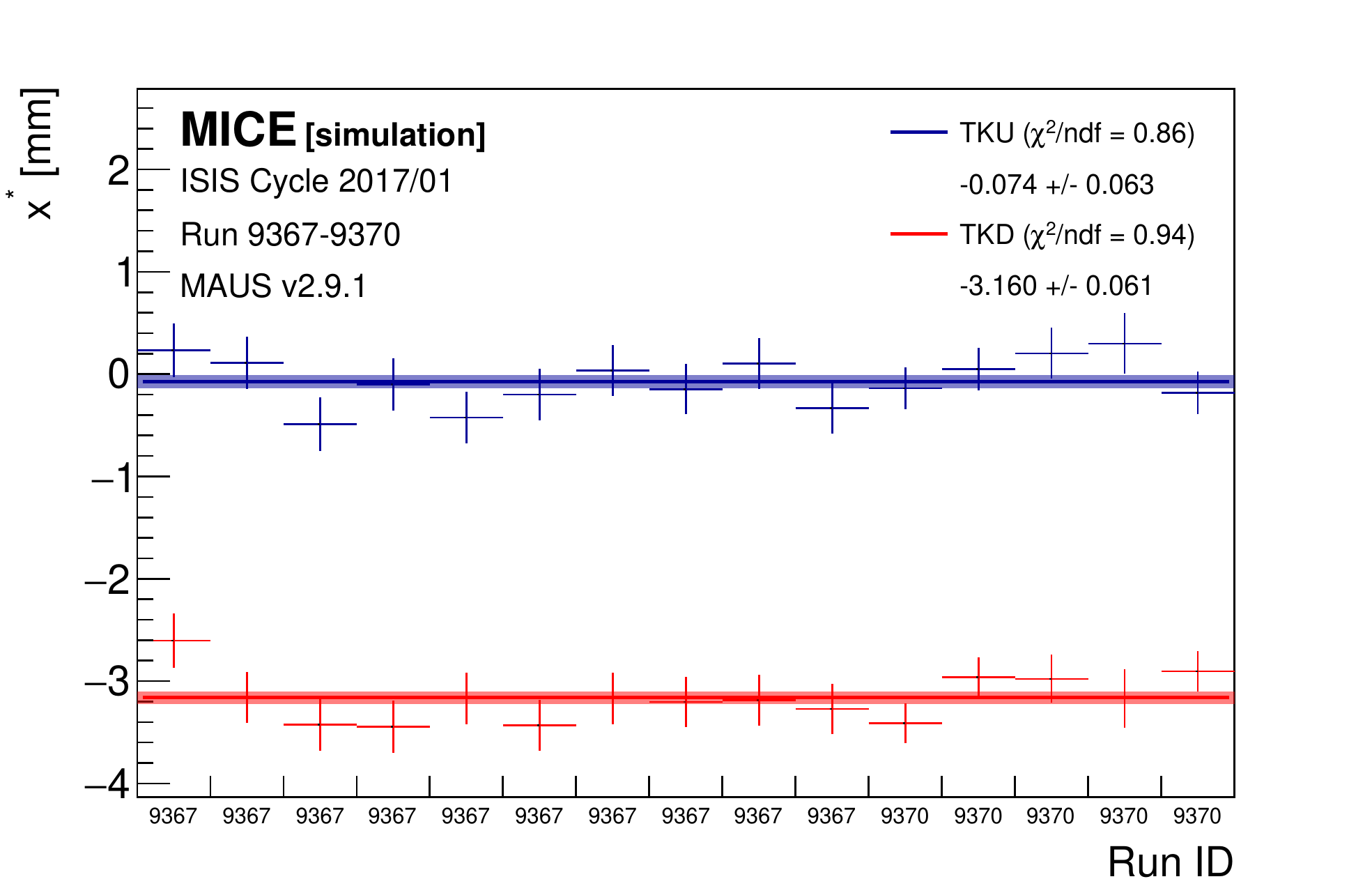}
	\end{minipage}
	\hfill
	\begin{minipage}[b]{.49\textwidth}
		\centering
		\includegraphics[width=\textwidth]{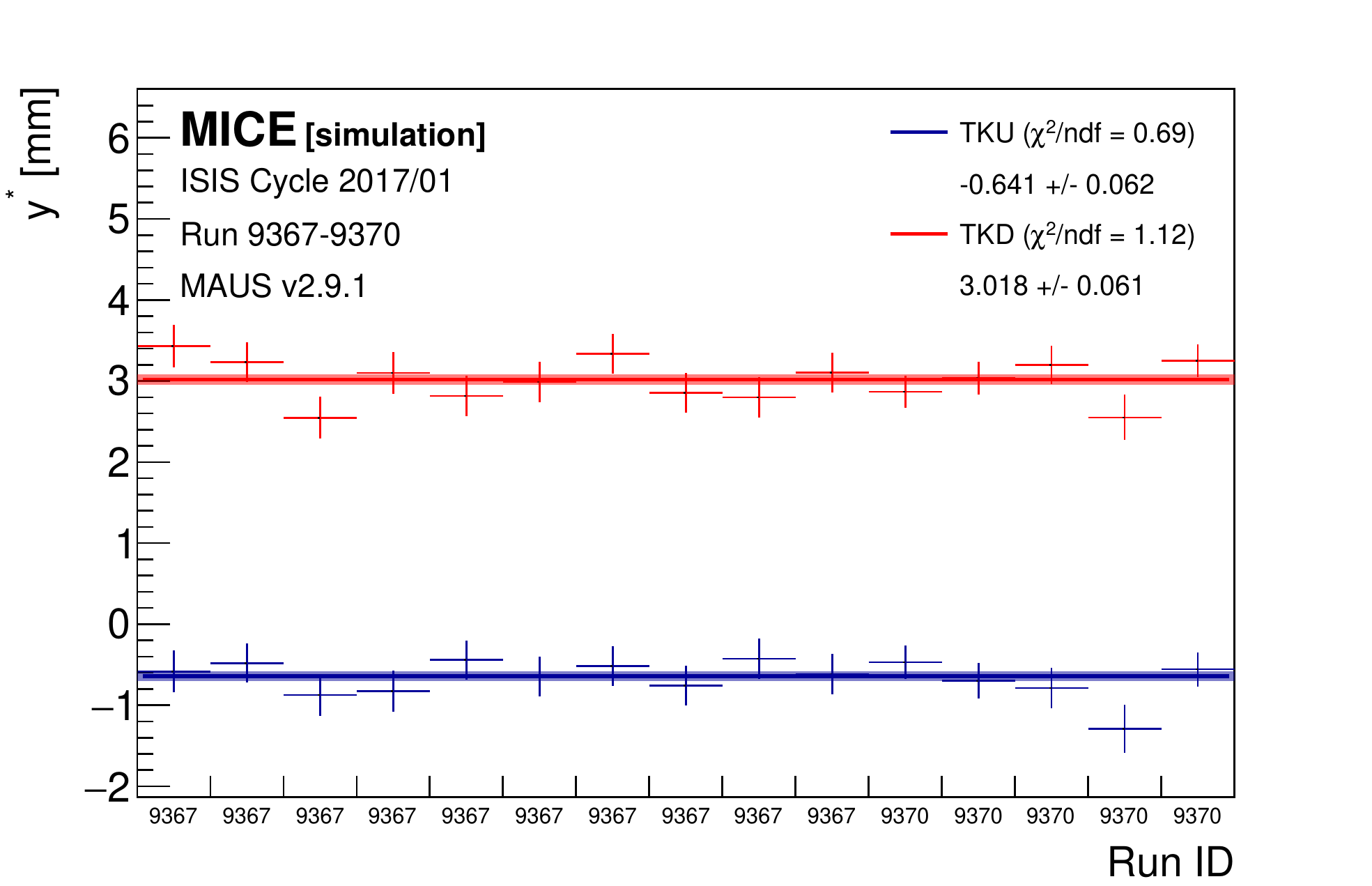}
	\end{minipage}
	
	\begin{minipage}[b]{.49\textwidth}
		\centering
		\includegraphics[width=\textwidth]{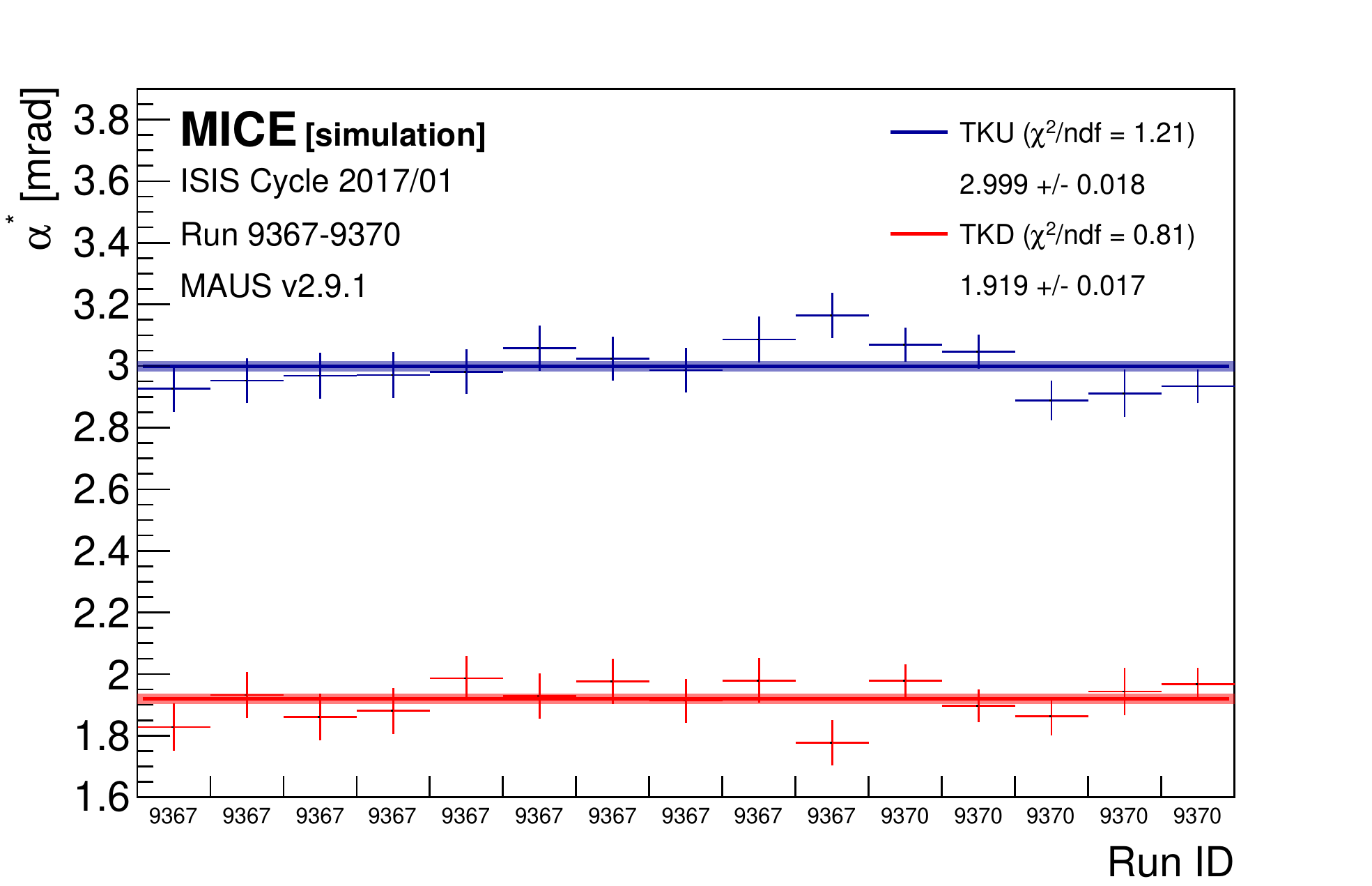}
	\end{minipage}
	\hfill
	\begin{minipage}[b]{.49\textwidth}
		\centering
		\includegraphics[width=\textwidth]{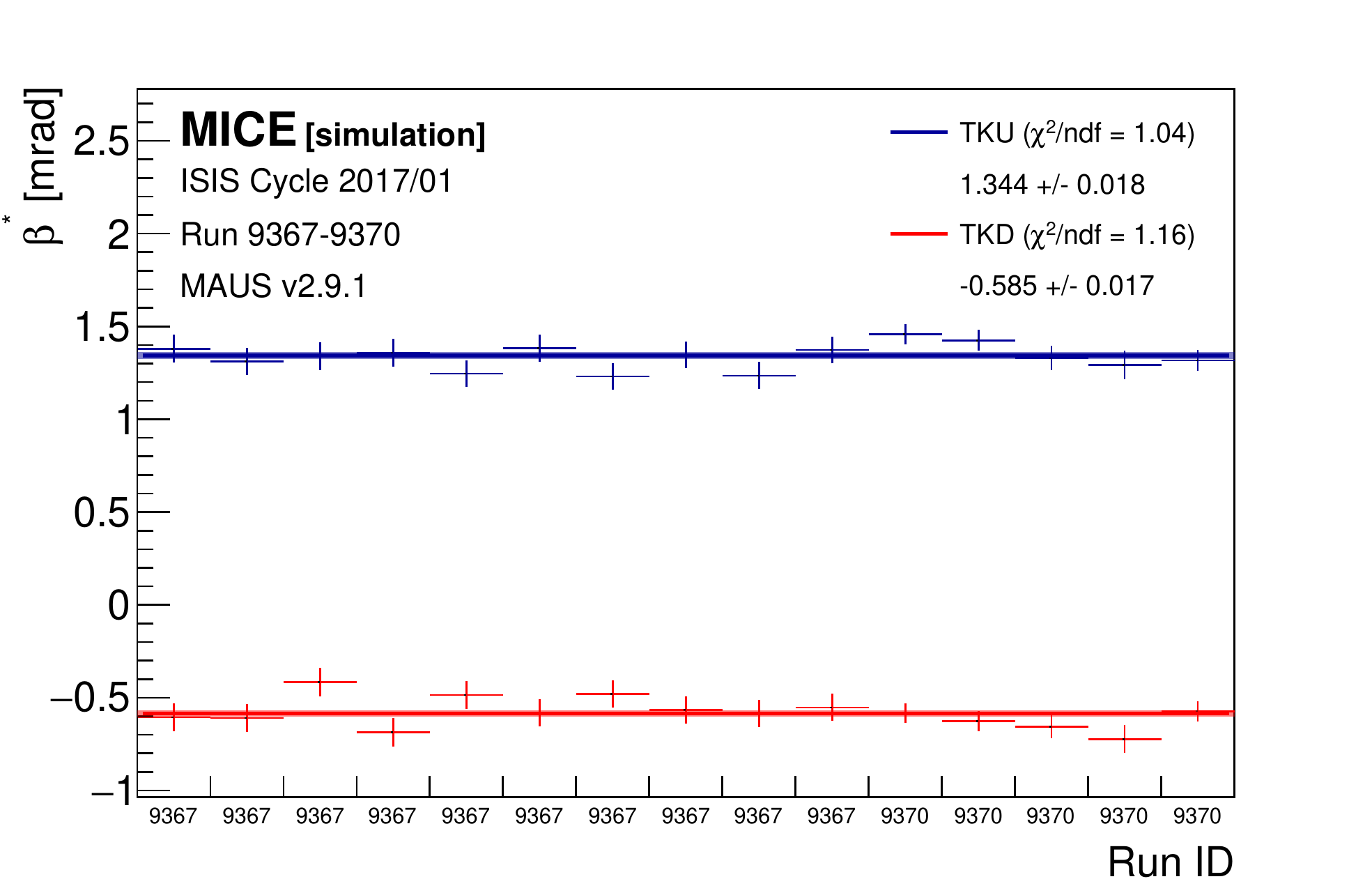}
	\end{minipage}
	\caption{Consistency of the alignment algorithm across MC runs produced for the 2017/01 ISIS user cycle.}
	\label{fig:runtorun}
\end{figure}

Each of the fifteen chunks described earlier were processed independently with the algorithm. Figure~\ref{fig:runtorun} compiles the alignment parameters measured for each run. The measurements are in good agreement with one another and show no significant discrepancy. The constant fit $\chi^2/\text{ndf}$ is close to unity for each fit, which indicates that there are no significant additional source of uncertainty. The optimal parameters are summarised in table\,\ref{tab:params}, compared to the true positions of the trackers in the Monte Carlo.

\begin{table}[h!]
	\centering
	\begin{tabular}{c|c|c|c|c}
		& \multicolumn{2}{c|}{TKU} & \multicolumn{2}{c}{TKD} \\
		\cline{2-5}
		& Truth & Best fit & Truth & Best fit \\
		\hline
		x [mm] & $-0.138$ & $-0.074\pm0.063$ & $-3.207$ & $-3.160\pm0.061$ \\
		y [mm] & $-0.689$ & $-0.641\pm0.062$ & $3.043$ & $3.018\pm0.061$\\
		$\alpha$ [mrad] & $3.014$ & $2.999\pm0.018$ & $1.919$ & $1.917\pm0.017$ \\
		$\beta$ [mrad] & $1.311$ & $1.344\pm0.018$ & $-0.568$ & $-0.585\pm0.017$
	\end{tabular}
	\caption{Summary table of the optimal alignment constants measured in the Monte Carlo.}
	\label{tab:params}
\end{table}

\subsection{Propagation}
The fitted parameters are used to yield the global track coordinates at the tracker $t=u,d$ centres, $(\xi_t,\upsilon_t,\zeta_t)$, through equation~\ref{eq:rot_matrix} and the global gradients $\xi'_t,\upsilon'_t$ through equation~\ref{eq:global_grad}. A corrected global track is propagated in an adjacent detector module $M$ at $\zeta_m$ through
\begin{equation}
\psi_t^m=\psi_t+\psi'_t(\zeta_m-\zeta_t),\,\psi=\xi,\upsilon.
\end{equation}

Provided exact corrections, a detector module $M$ that measures a global position $(\xi_m,\upsilon_m,\zeta_m)$ verifies
\begin{equation}
 \left\{
 \begin{array}{l}
  \langle \psi_m-\psi_t^m\rangle=0 \\
  \langle \psi'_m-\psi'_t\rangle=0
 \end{array}
 \right.,\,\psi=\xi,\upsilon.
\end{equation}

\begin{figure} [!htb]
\centering
\begin{minipage}[b]{.475\textwidth}
\centering
\includegraphics[width=\textwidth]{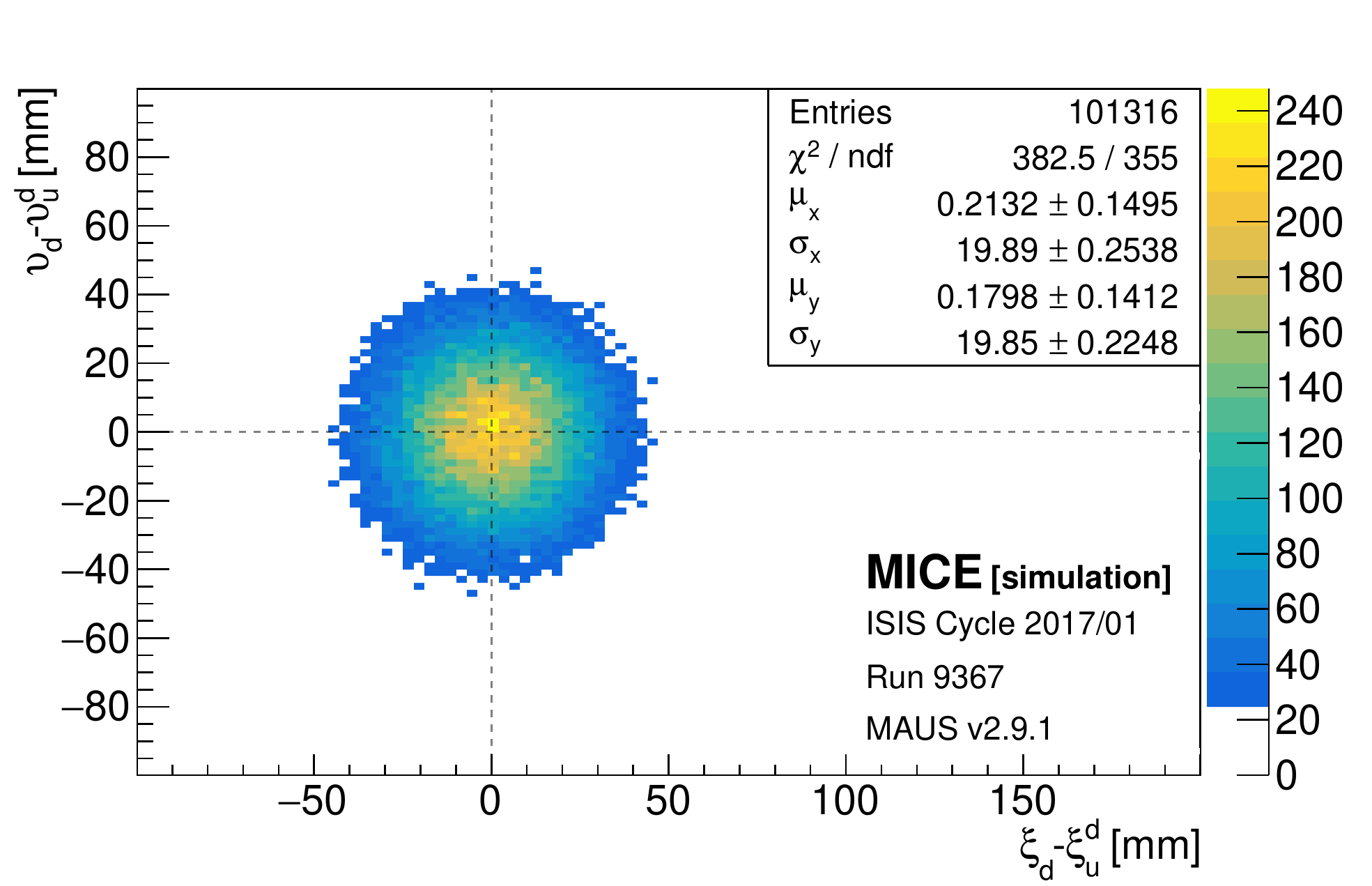}
\end{minipage}
\hfill
\begin{minipage}[b]{.475\textwidth}
\centering
\includegraphics[width=\textwidth]{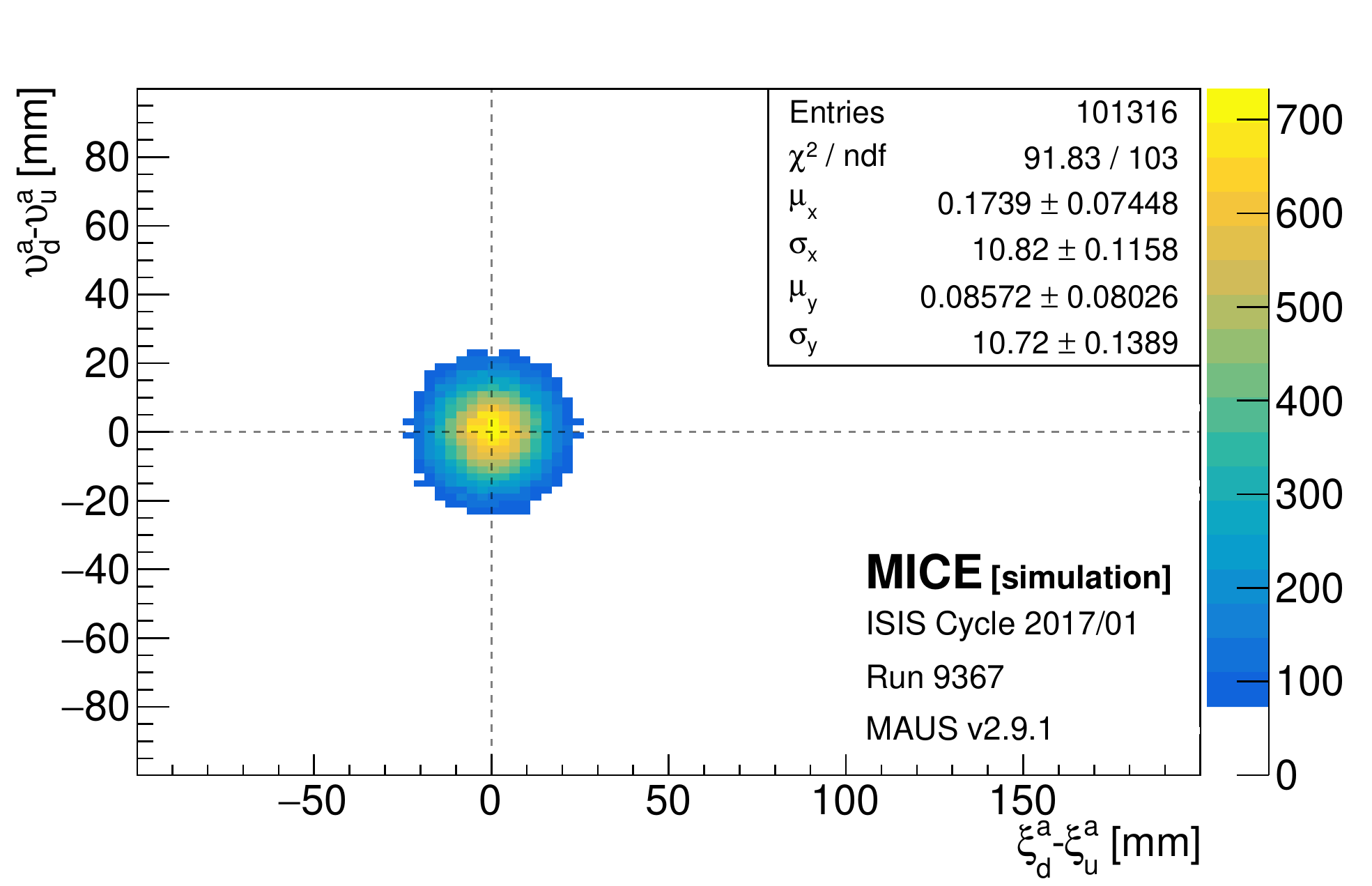}
\end{minipage}

\begin{minipage}[b]{.475\textwidth}
\centering
\includegraphics[width=\textwidth]{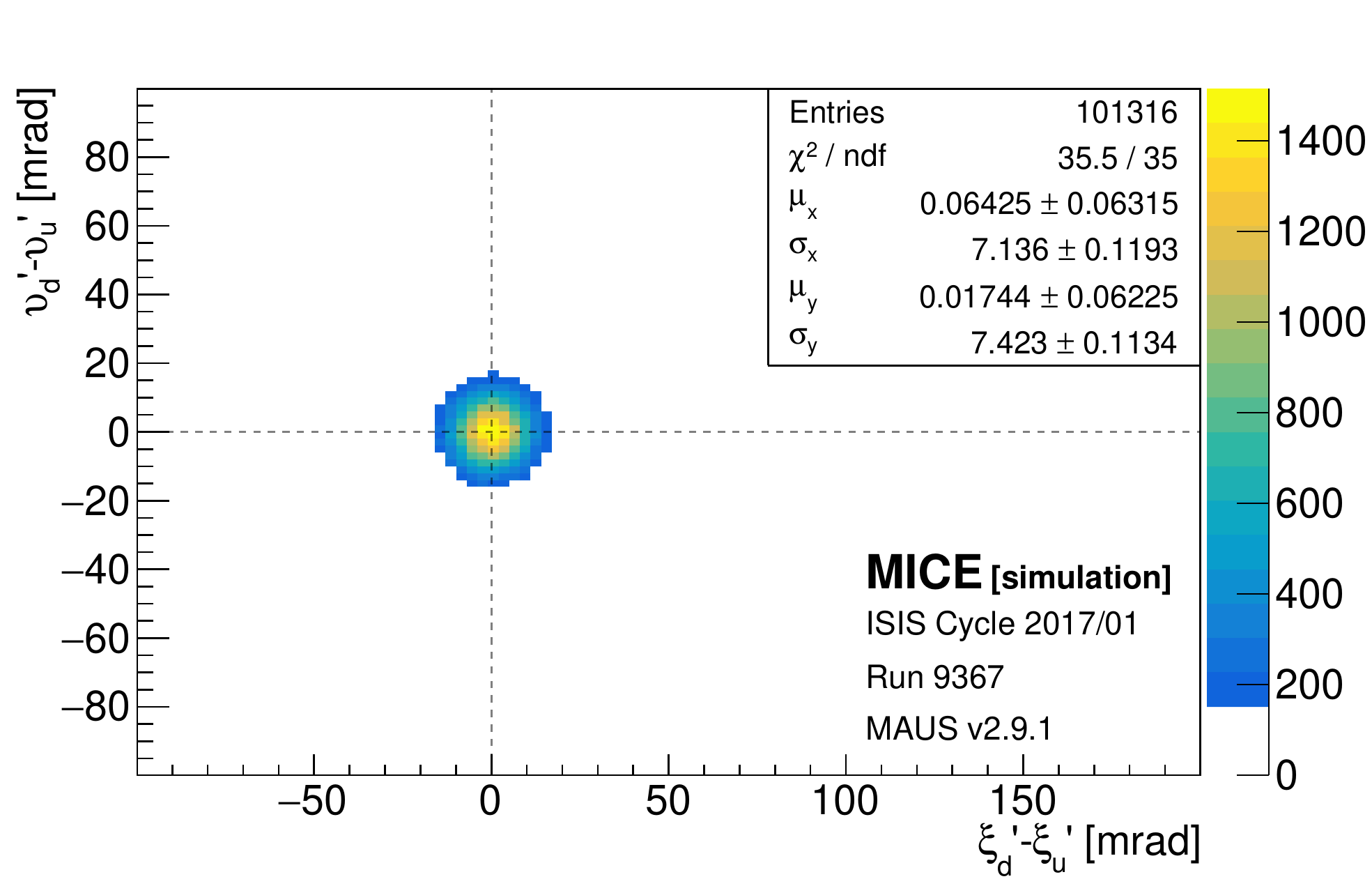}
\end{minipage}
\hfill
\begin{minipage}[b]{.475\textwidth}
\centering
\includegraphics[width=\textwidth]{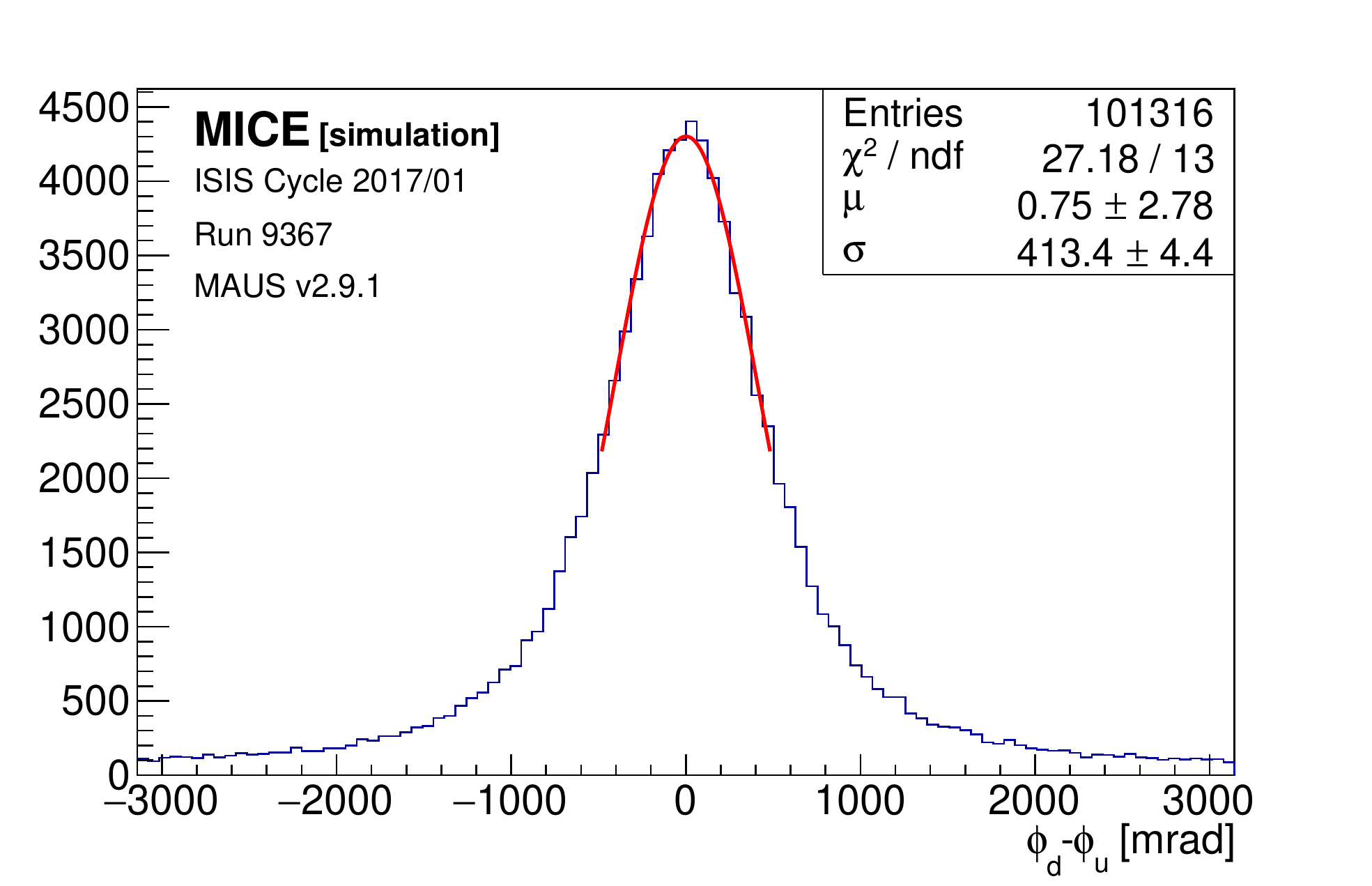}
\end{minipage}
\caption{Tracker-to-tracker residual distributions in position (\textbf{top}) and angle (\textbf{bottom}).}
\label{fig:tracker_residuals}
\end{figure}

As a consistency check, the tracks are first propagated between the two trackers. The selection described in section\,\ref{sec:criterion} is applied but no compensation is added for lost tracks. The results are shown in figure~\ref{fig:tracker_residuals}. The top left and right distributions show the residuals between the TKU and TKD tracks at the centre of the downstream tracker and at the level of the absorber, respectively. The bottom two histograms show the agreement between the angles measured upstream and downstream. The azimuthal angle residuals show consistency between the roll of the two trackers. The results are compatible with what is expected from aligned trackers.

\begin{figure} [!htb]
\centering
\begin{minipage}[b]{.475\textwidth}
\centering
\includegraphics[width=\textwidth]{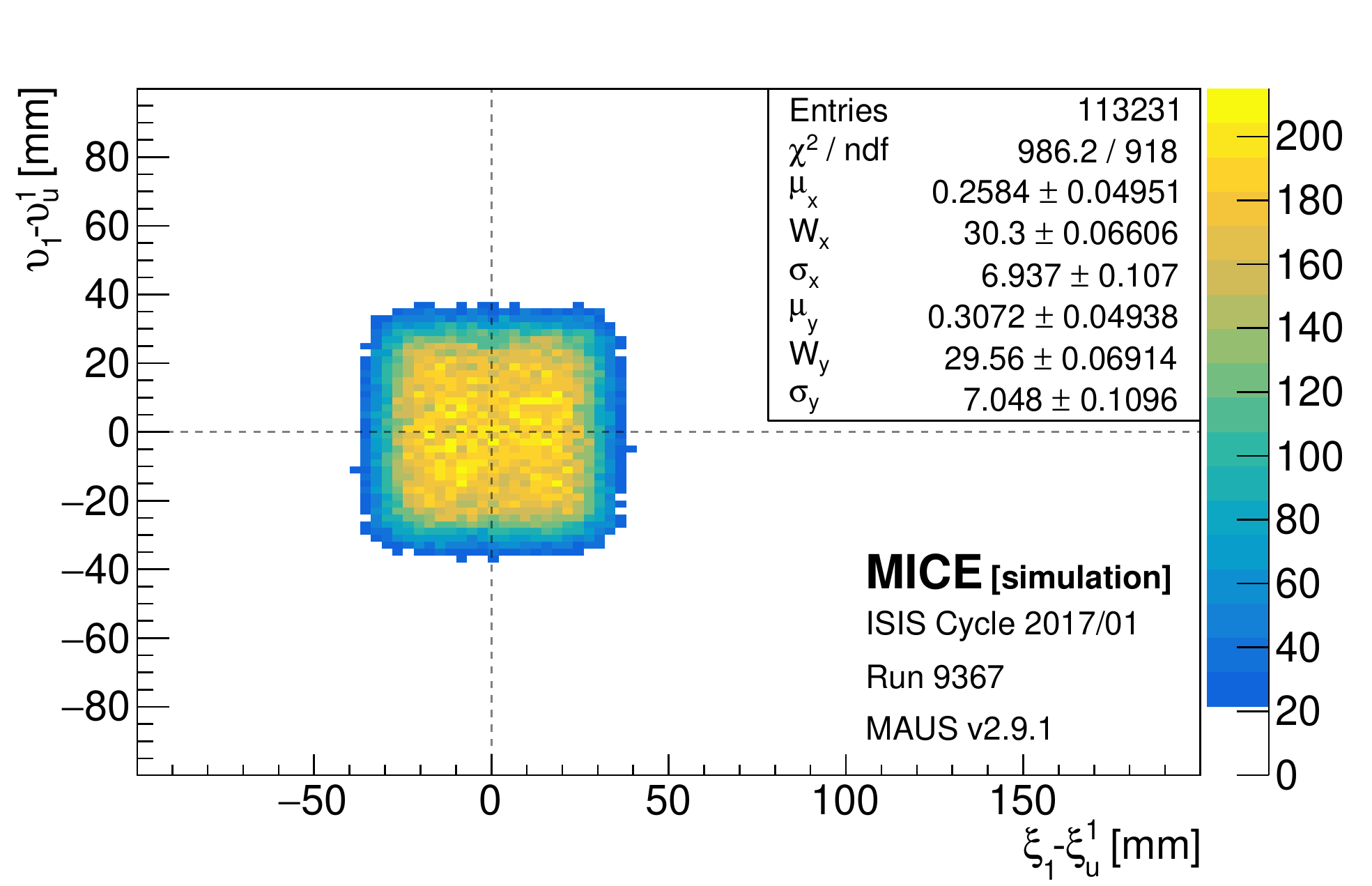}
\end{minipage}
\hfill
\begin{minipage}[b]{.475\textwidth}
\centering
\includegraphics[width=\textwidth]{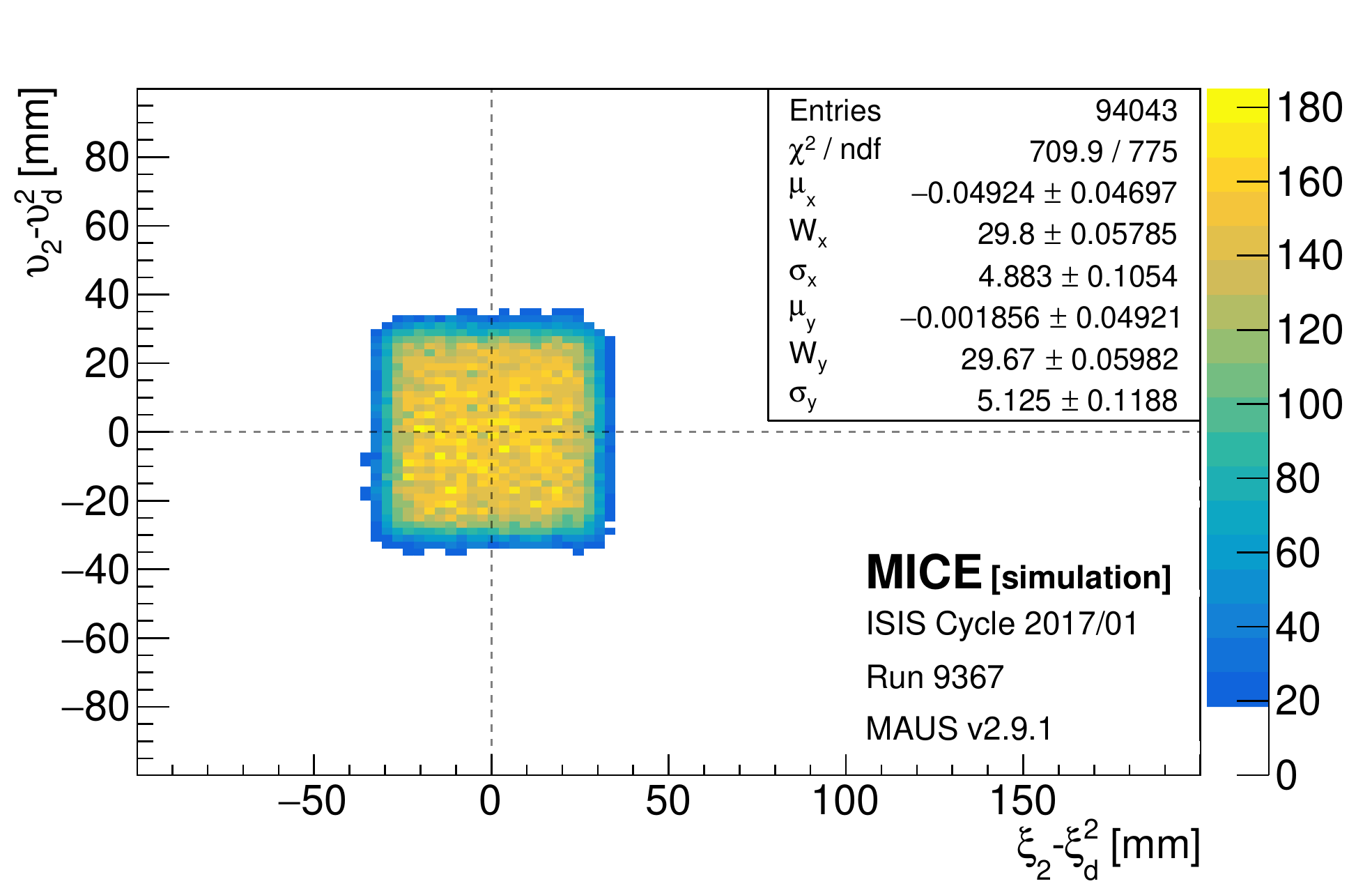}
\end{minipage}

\begin{minipage}[b]{.475\textwidth}
\centering
\includegraphics[width=\textwidth]{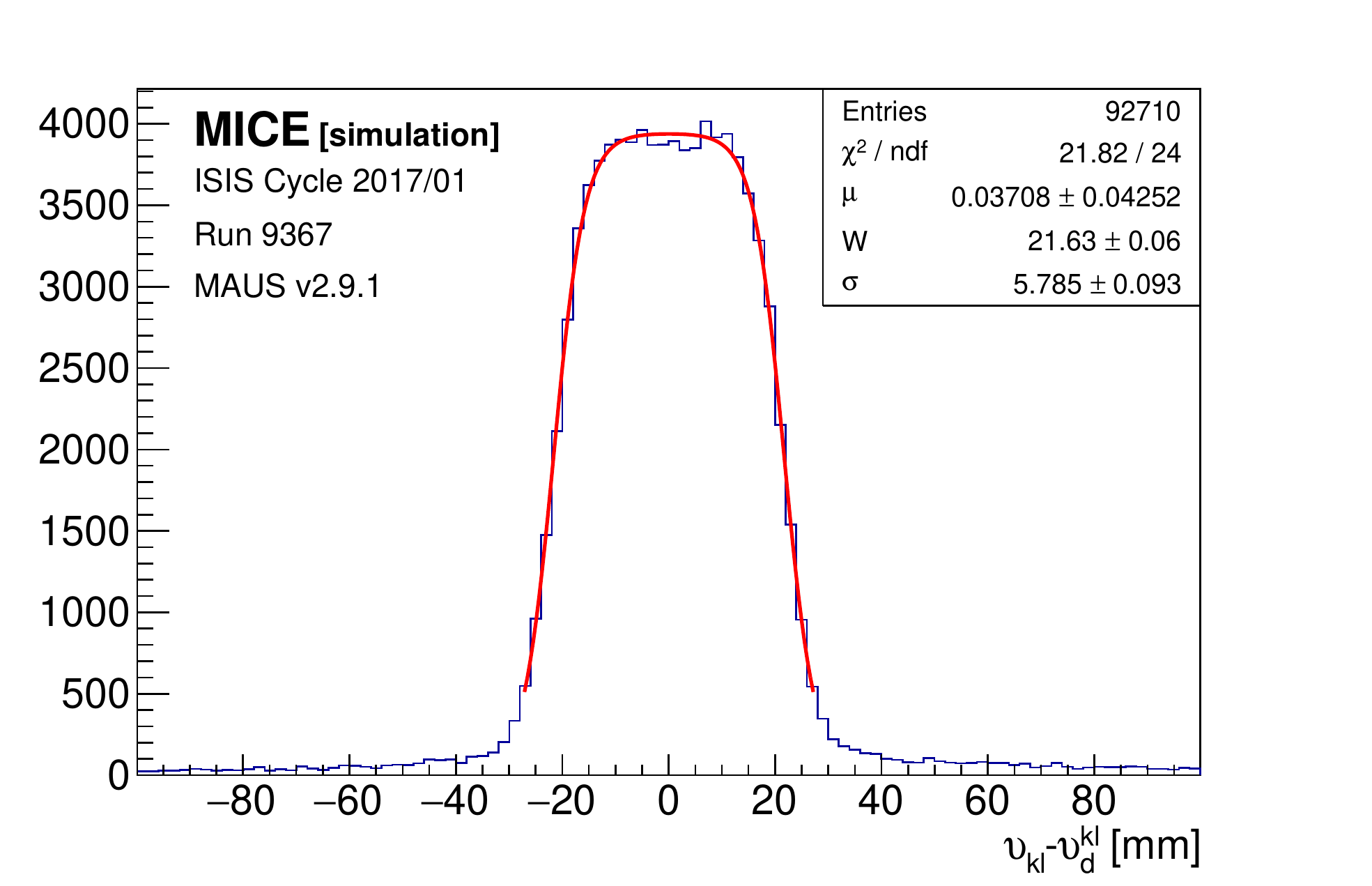}
\end{minipage}
\hfill
\begin{minipage}[b]{.475\textwidth}
\centering
\includegraphics[width=\textwidth]{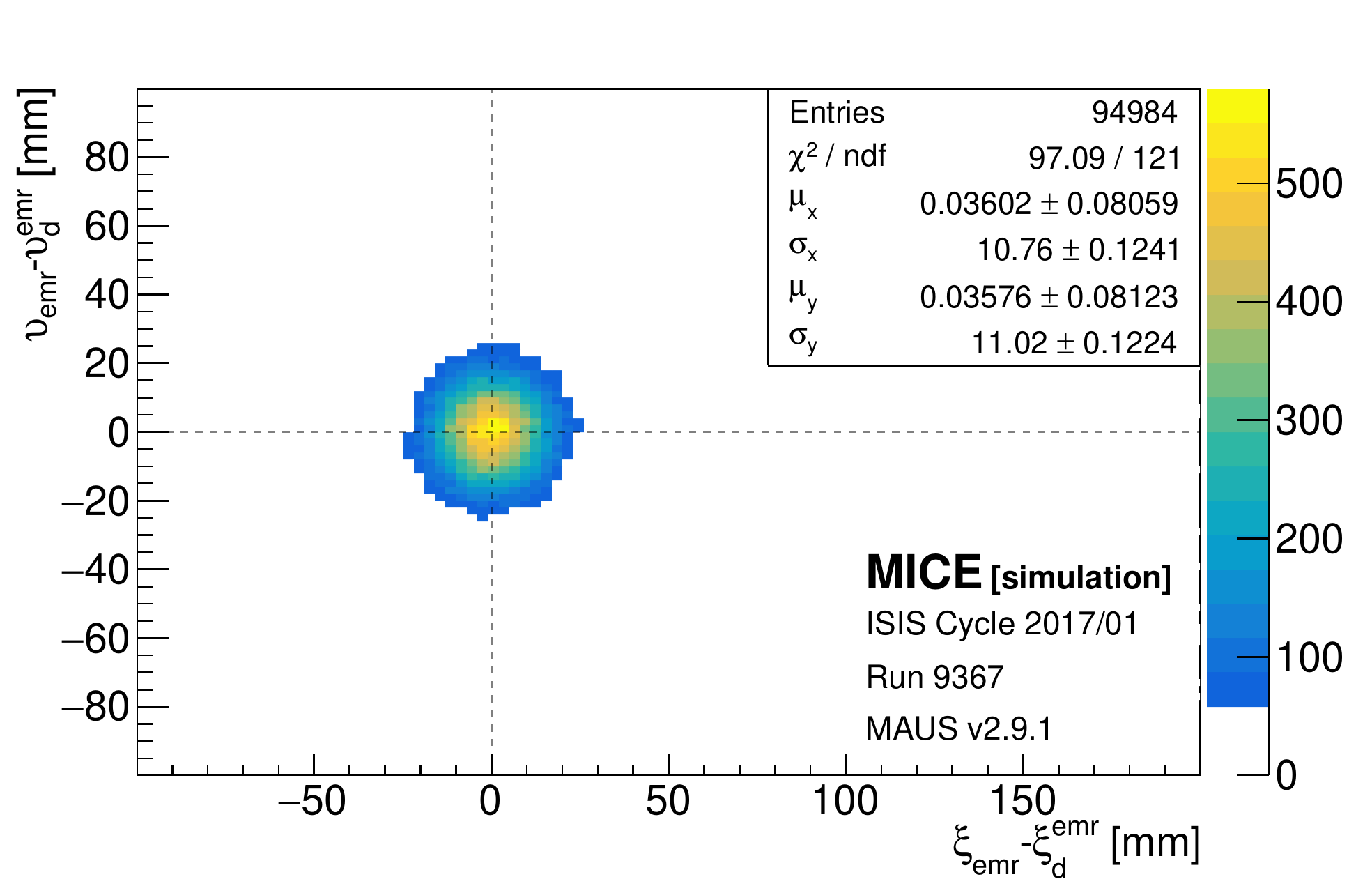}
\end{minipage}
\caption{Tracker to particle identification detectors residual distributions (TOF1, TOF2, KL and EMR).}
\label{fig:pid_residuals}
\end{figure}

The upstream tracker tracks are extrapolated into TOF1 and the downstream tracker tracks are propagated into the three downstream particle identification detectors: TOF2, the KL and the EMR. The residual plots are represented in figure~\ref{fig:pid_residuals}. The values obtained show good agreement between the tracks and the space points measured in other MICE detectors.

Special care is taken when evaluating the central value of the residual distributions. The two trackers and the Electron-Muon Ranger have a sufficient spacial resolution to follow a near-Gaussian distribution. The residuals involving these detectors are fitted with a standard multivariate normal of mean $\bm{\mu}$ and width $\bm{\sigma}$ between the two half-maximum, i.e. in the range $\bm{\mu}\pm1.1775\,\bm{\sigma}$. The TOF hodoscopes and the KL do not have a sufficient resolution to produce residuals that follow a Gaussian distribution. A probability density function of the form
\begin{equation}
h(x)=\frac{1}{4W}\left(\tanh\left[\frac{x-\mu+W}{\sigma}\right]-\tanh\left[\frac{x-\mu-W}{\sigma}\right]\right)
\end{equation}
is used in each projection to fit the residuals involving the low granularity detectors. The constant $\mu$ represents the central value of the residual distribution, $\sigma$ the residual width and $W$ the half-width of one of the low-resolution detector pixel. The parameters obtained for each of the fits are represented in table\,\ref{tab:res}. The values found for $W$ are consistent with pixels of 6\,cm in TOF1 and TOF2 and of 4.4\,cm in the KL.

\begin{table}[!htb]
	\centering
	\begin{tabular}{c|c|c|c|c|c}
		Origin & Target & Variable & $\mu$ [mm]/[mrad] & $\sigma$ [mm]/[mrad] & $W$ [mm]\\
		\hline
		\multirow{7}{*}{TKU} & \multirow{2}{*}{TOF1} & $\xi$ & $0.258\pm0.050$ & $6.937\pm0.107$ & $30.30\pm0.07$ \\
		& & $\upsilon$ & $0.3072\pm0.049$ & $7.048\pm0.110$ & $29.56\pm0.07$ \\
		\cline{2-6}
		& \multirow{5}{*}{TKD} & $\xi$ & $0.213\pm0.150$ & $19.890\pm0.254$ & -- \\
		& & $\upsilon$ & $0.180\pm0.141$ & $19.850\pm0.225$ & --\\
		& & $\xi'$ & $0.064\pm0.063$ & $7.136\pm0.119$ & --\\
		& & $\upsilon'$ & $0.017\pm0.062$ & $7.423\pm0.113$ & --\\
		& & $\phi$ & $0.75\pm2.78$ & $413.4\pm4.4$ & -- \\
		\hline
		\multirow{5}{*}{TKD} & \multirow{2}{*}{TOF2} & $\xi$ & $-0.049\pm0.047$ & $4.883\pm0.105$ & $29.80\pm0.06$ \\
		& & $\upsilon$ & $-0.002\pm0.049$ & $5.125\pm0.119$ & $29.67\pm0.06$ \\
		\cline{2-6}
		& KL & $\upsilon$ & $0.037\pm0.042$ & $5.785\pm0.093$ & $21.63\pm0.06$ \\
		\cline{2-6}
		& \multirow{2}{*}{EMR} & $\xi$ & $0.036\pm0.081$ & $10.760\pm0.124$ & -- \\
		& & $\upsilon$ & $0.036\pm0.081$ & $11.020\pm0.122$ & --
	\end{tabular}
	\caption{Residual distributions fitting parameters.}
	\label{tab:res}
\end{table}

\section{Conclusions}
The beam-based detector alignment analysis has shown that all of the detectors in the MICE beam line reconstruct space points and tracks in a consistent fashion. Due to the embedded nature of the scintillating fibre trackers, it is necessary to use straight particle tracks to probe their position within the superconducting magnets bore. The adjustment is achieved by using detectors whose positions are well known from laser telemetry surveys as reference points to measure the offsets in the two trackers.

The alignment algorithm was tested on two large, highly differentiated Monte Carlo samples duplicating runs 9367 and 9370 taken during the 2017/01 ISIS user cycle. The alignment constants measured are in excellent agreement with the true tracker positions. The resolution on the alignment constants improves as $2\sigma_W/\sqrt{N}$ with $\sigma_W\simeq3$\,mrad in angle and $\sigma_W\simeq10$\,mm in position.

\clearpage
\bibliographystyle{utphys}
\bibliography{bibliography}

\end{document}